\documentclass[prb,superscriptaddress,twocolumn,floatfix,letterpaper]{revtex4}

\usepackage{amssymb, amsmath}
\usepackage{mathrsfs}
\usepackage[usenames]{color}
\usepackage{subfigure, wrapfig}
\usepackage[dvips]{graphicx}
\usepackage[normalem]{ulem}

\newcommand{\nquad}{\!\!\!\!\!}
\newcommand{\txt}[1]{\textnormal{#1}}
\newcommand{\punct}[1]{\textrm{ #1}}
\newcommand{\etal}{{\itshape et al.\@}}

\newcommand{\set}[1]{\{ #1 \}}
\newcommand{\vv}[1]{\mathbf{#1}}
\newcommand{\vvsym}[1]{\boldsymbol{#1}}
\newcommand{\uv}[1]{\hat{#1}}

\newcommand{\dg}{\dagger}
\newcommand{\up}{\uparrow}
\newcommand{\dn}{\downarrow}
\newcommand{\ket}[1]{| #1 \rangle}
\newcommand{\bra}[1]{\langle #1 |}

\newcommand{\avg}[1]{\langle #1 \rangle}
\newcommand{\op}[1]{#1}
\newcommand{\opdag}[1]{#1^\dg}

\newcommand{\E}{\mathcal{E}} 
 
\newcommand{\bigO}{\mathcal{O}}

\newcommand{\tr}{\operatorname{tr}}
\newcommand{\cc}[2]{(c^\dg_{#1} c_{#2})}
\newcommand{\ff}[2]{(f^\dg_{#1} f_{#2})}
\newcommand{\SdotS}[2]{\vv{S}_{#1} \cdot \vv{S}_{#2}}
\newcommand{\SSS}[3]{\mathcal{S}_{#1;#2;#3}}
\newcommand{\tchi}{\tilde{\chi}}
\newcommand{\Hb}{\textrm{Hb}}
\newcommand{\Hbb}{\textrm{(Hb)}}
\newcommand{\DOS}{\mathcal{D}}
\newcommand{\tris}{\triangle\textrm{'s}}
\newcommand{\aone}{\vv{a}_1}
\newcommand{\atwo}{\vv{a}_2}
\newcommand{\pol}{\vvsym{\epsilon}}
\newcommand{\Imag}{\operatorname{Im}}

\usepackage{pstricks}
\usepackage{pst-plot}
\usepackage{pstricks-add}

\input{symbols.pst}
\input{main_graphics.pst}
\input{appendix_graphics.pst}

\usepackage[dvips]{hyperref}

\begin{document}

\title{Raman signature of the U(1) Dirac spin-liquid state in spin-1/2 kagome system}

\author{Wing-Ho Ko}
\affiliation{Department of Physics, Massachusetts Institute of Technology, Cambridge, Massachusetts, 02139, USA}
\author{Zheng-Xin Liu}
\author{Tai-Kai Ng}
\affiliation{Department of Physics, Hong Kong University of Science and Technology, Hong Kong, People's Republic of China}
\author{Patrick A. Lee}
\affiliation{Department of Physics, Massachusetts Institute of Technology, Cambridge, Massachusetts, 02139, USA}

\date{November 3, 2009}

\begin{abstract} 
We followed the Shastry--Shraiman formulation of Raman scattering in Hubbard systems and considered the Raman intensity profile in the spin-1/2 ``perfect'' kagome lattice herbertsmithite ZnCu$_3$(OH)$_6$Cl$_2$, assuming the ground state is well-described by the U(1) Dirac spin-liquid state. In the derivation of the Raman $T$-matrix, we found that the spin-chirality term appears in the $A_{2g}$ channel in the kagome lattice at the $t^4/(\omega_i-U)^3$ order, but (contrary to the claims by Shastry and Shraiman) vanishes in the square lattice to that order. In the ensuing calculations on the spin-1/2 kagome lattice, we found that  the Raman intensity profile in the $E_g$ channel is invariant under an arbitrary rotation in the kagome plane, and that in all ($A_{1g}$, $E_g$, and $A_{2g}$) symmetry channels the Raman intensity profile contains broad continua that display power-law behaviors at low energy, with exponent approximately equal to 1 in the $A_{2g}$ channel and exponent approximately equal to 3 in the $E_g$ and the $A_{1g}$ channels. For the $A_{2g}$ channel, the Raman profile also contains a characteristic $1/\omega$ singularity, which arose in our model from an excitation of the emergent U(1) gauge field.
\end{abstract}

\maketitle 

\section{Introduction} \label{sect:intro} 

Quantum spin liquids, which are quantum ground states of interacting spin systems that break no symmetries, have long been a fascination among the theoretical condensed matter community. After years of experimental searches, several promising candidates finally emerged, including the ``perfect'' spin-1/2 kagome lattice herbertsmithite ZnCu$_3$(OH)$_6$Cl$_2$, which shows no signs of magnetic ordering down to a temperature of 50~mK, despite having a nearest-neighbor antiferromagnetic exchange $J \approx 190$~K.\cite{Helton:PRL:2007, Mendels:PRL:2007, Ofer:arXiv:2006}

Ever since the successful synthesis of herbertsmithite,\cite{Shores:JACS:2005} a host of experimental techniques have been applied to study the material, including thermodynamic measurements,\cite{Helton:PRL:2007, Ofer:arXiv:2006, Vries:PRL:2008, Bert:PRB:2007} neutron diffraction,\cite{Helton:PRL:2007, Lee:NM:2007} NMR,\cite{Ofer:arXiv:2006,
Olariu:PRL:2008, Imai:PRL:2008} and $\mu$SR.\cite{Ofer:arXiv:2006, Mendels:PRL:2007} Unfortunately, the  experimental results accumulated thus far are still insufficient to determine if the material is truly a quantum spin liquid. In particular, the valence bond solid (VBS) state proposed in Ref.~\onlinecite{Hastings:PRB:2001}~and~\onlinecite{Nikolic:PRB:2003} remains a possible alternative to the U(1) Dirac spin-liquid (DSL) state proposed in Ref.~\onlinecite{Ran:PRL:2007}. In order to settle the debate of which theoretical model best describe the quantum state in herbertsmithite, further experimental probes, with guidances from theory, are probably required.

Recently, Cepas \etal\cite{Cepas:PRB:2008} considered Raman scattering on the spin-1/2 kagome system and concluded that a generic spin-liquid state can be distinguished from a generic VBS state by the polarization dependence of the signal. They also obtained a more detailed prediction of the Raman intensity using a random phase approximation, which may be too crude given the subtle orders\cite{Hermele:PRB:2008} that may be present in the system.

In this paper, we consider the specific scenario that herbertsmithite is described by the DSL state, and make theoretical predictions of the experimental signatures that should be present in the Raman scattering. In contrast to the predictions by Cepas \etal\ (who argued that signals consist of several sharp Lorentzian peaks), we predict that \emph{broad continua should be present in all ($E_g$, $A_{1g}$, and $A_{2g}$) symmetry channels}, with a super-linear power law at low energy in the $A_{1g}$ and the $E_g$ channels, and a linear power law in at low energy in the $A_{2g}$ channel. In addition, we predict that \emph{a $1/\omega$ singularity should be observed in the $A_{2g}$ channel.} From a theoretical perspective, this $1/\omega$ singularity is particularly interesting, since it can be thought of as arising from an excitation associated with an emergent gauge boson in the system.

The paper is organized as follows: In Sec.~\ref{sect:SS_formulation}, the Shastry--Shraiman formulation of Raman scattering in Hubbard systems is reviewed and our disagreement with the original results by Shastry and Shraiman on the square lattice is discussed, after which the relevant results for the kagome lattice is presented (the detailed derivations and further discussions are relegated to Appendices).  In Sec.~\ref{sect:spin_liquid}, the U(1) Dirac spin-liquid model is reviewed, with important features in the mean-field theory highlighted. The Raman intensity profile in $E_g$ channel is then presented in Sec.~\ref{sect:Eg}, and the analogous results for the $A_{1g}$ and $A_{2g}$ channel are presented in Secs.~\ref{sect:A1g}~and~\ref{sect:A2g}, respectively. Further discussions on these results are presented in Sec.~\ref{sect:discuss}.

\section{Shastry--Shraiman formulation} \label{sect:SS_formulation}

Being a strongly correlated material, herbertsmithite can be described using a one-band Hubbard model:
\begin{equation} \label{eq:H_Hb}
H_{\Hb} = H_t + H_U = - \sum_{ij,\sigma} t_{ij} c^\dg_{i\sigma} c_{j\sigma} + U \sum_{i} n_{i\up} n_{i\dn} \punct{,}
\end{equation}
where $i$, $j$ label lattice sites and $\sigma = \up, \dn$ labels spin. $c^\dg_i$ ($c_i$) is the electron creation (annihilation) operator on site $i$, and $n_{i\sigma} = c^\dg_{i\sigma} c_{i\sigma}$.

Coupling to the external electromagnetic field can be incorporated by the replacement  $c^\dg_{i\sigma} c_{j\sigma} \mapsto c^\dg_{i\sigma} c_{j\sigma} \exp(\frac{ie}{\hbar c}\int_{j}^{i} \vv{A}\cdot d\vv{x})$. Expanding this exponential and including also the free photon Hamiltonian $H_{\gamma}$, the Hamiltonian now reads:
\begin{equation} \label{eq:H}
\begin{aligned}
H & = H_{\Hb} + H_{\gamma} + H_{C} \punct{,}\\
H_{C} & = -\sum_{ij,\sigma} t_{ij} c^\dg_{i\sigma} c_{j\sigma} 
	\bigg( \frac{ie}{\hbar c} \vv{A}(\frac{\vv{x}_i+\vv{x}_j}{2}) \cdot (\vv{x}_i - \vv{x}_j) \\
	& \qquad - \frac{e^2}{\hbar^2 c^2} \Big( \vv{A}(\frac{\vv{x}_i+\vv{x}_j}{2}) \cdot (\vv{x}_i - \vv{x}_j) \Big)^2	+  \cdots \bigg) \punct{,}\\ 
H_{\gamma} & = \sum_{\vv{q}} \omega_{\vv{q}} a^{\alpha\dg}_{\vv{q}} a^\alpha_{\vv{q}} \punct{,}
\end{aligned}
\end{equation}
where $a^{\alpha\dg}_\vv{q}$ ($a^\alpha_\vv{q}$) denotes the photon creation (annihilation) operator at momentum $\vv{q}$ and polarization $\alpha$, and $\vv{A}(\vv{x})$ denotes the photon operator in real space. The $\cdots$ are terms at higher order in $\vv{A}$. 

By treating $H_{C}$ as a time-dependent perturbation, the transition rate from an initial state $\ket{i}$ to a final state $\ket{f}$ is given by: 
\begin{equation} \label{eq:rate}
\Gamma_{fi} = 2\pi |\bra{f} T \ket{i}|^2 \delta(\E_f - \E_i) \punct{,}
\end{equation}
where $\E_i$ ($\E_f$) is the energy of the initial (final) state and $T = H_{C} + H_{C} (\E_i - H_{\txt{Hb}} - H_\gamma + i \eta)^{-1} H_{C} +\cdots $ is the $T$-matrix.

Since the fine-structure constant $e^2/\hbar c \approx 1/137$ is small and since we are interested in Raman processes (one photon in, one photon out), only terms second order in $\vv{A}$ need to be retained. At this order, the $T$-matrix reads:
\begin{equation} \label{eq:Tmatrix}
T = H^{(2)}_{C} + H^{(1)}_{C} \frac{1}{\E_i - (H_{\Hb} + H_{\gamma}) + i \eta} H^{(1)}_{C}
= T_{\txt{NR}} + T_{\txt{R}} \punct{,}
\end{equation}
where $H^{(n)}_{C}$ denotes the part of $H_{C}$ that is $n$-th order in $\vv{A}$. The subscript R and NR on the last equality stands for resonant and non-resonant, respectively.

We are interested in a half-filled system ($\avg{\sum_{\sigma} n_{i\sigma}} = 1$) in the localized regime ($U \gg t$), in which both the initial and the final state belongs to the near-degenerate ground-state manifold $n_{i\up}n_{i\dn}=0$. In such case, $T_{\txt{NR}}$ has no matrix element that directly connects between the initial and the final states. Hence, only $T_{\txt{R}}$ is relevant for our purpose. 

Let $\omega_i$ ($\omega_f$), $\vv{k}_i$ ($\vv{k}_f$), and $\vv{e}_i$ ($\vv{e}_f$) be the frequency, momentum, and polarization of the incoming (outgoing) photon, respectively. Then, $\E_i = \omega_i + \E^{\Hbb}_i = \omega_i + \bigO(t^2/U)$, where $H_{Hb} \ket{i} = \E^{\Hbb}_i \ket{i}$; and $\vv{A}(\vv{x}) \mapsto g_i \vv{e}_i a^{e_i}_{\vv{k}_i} e^{i \vv{k}_i \cdot \vv{x}} + g_f \bar{\vv{e}}_{f} a^{e_f\dg}_{\vv{k}_f} e^{-i \vv{k}_f \cdot \vv{x}}$, where $g_i = \sqrt{h c^2/\omega_{\vv{k}_i} \Omega}$ and $g_f = \sqrt{h c^2/\omega_{\vv{k}_f} \Omega}$, with $\Omega$ being the appropriate volume determined by the size of the sample and/or the size of the laser spot. In much of the following we shall assume as typical that the momenta carried by the photons are much smaller than the inverse lattice spacing, and hence $e^{-i \vv{k}_i \cdot \vv{x}} \approx e^{-i \vv{k}_f \cdot \vv{x}} \approx 1$. We shall also assume that the system is near resonance, so that $U \gg |\omega_i - U| \gtrsim |t|$. Consequently, henceforth we shall keep only terms that are zeroth order in $t/U$ and expand in powers of $t/(\omega_i-U)$.

Since the initial and final states both belong to the near-degenerate ground-state manifold, it should be possible to re-express $T_{\txt{R}}$ in terms of spin operators. A procedure for doing so was developed by Shastry and Shraiman.\cite{Shastry:PRL:1990, Shastry:IJMPB:1991} A first step in the derivation is to expand the denominator of $T_{\txt{R}}$:

\begin{widetext} 
\begin{equation} \label{eq:expand}
T_\txt{R} = H^{(1)}_{C} \frac{1}{\E_i - (H_{\Hb} + H_{\gamma}) + i \eta} H^{(1)}_{C} = 
H^{(1)}_{C} \frac{1}{\E_i - H_U - H_\gamma + i\eta} \sum_{n=0}^\infty \left( H_t \frac{1}{\E_i - H_U - H_\gamma + i\eta} \right)^n H^{(1)}_{C} \punct{.}
\end{equation}
\end{widetext}

Next, a spin quantization axis is fixed and the initial states $\ket{i} = \ket{\set{\sigma}} \otimes \ket{\vv{k}_i,\vv{e}_i}$ and final states $\ket{f} = \ket{\set{\sigma'}} \otimes \ket{\vv{k}_f,\vv{e}_f}$ are taken to be a direct product of a definite spin state in position basis with a photon energy eigenstate.\footnote{This introduces a small nuance that \protect{$\E_i$} can no-longer be treated as a scaler but must be considered as a matrix that depends on the initial and final spin states (but independent of the intermediate states). However, the off-diagonal terms of this matrix is of order \protect{$t/U$} and hence negligible.} Then, a complete set of states is inserted in between the operators in Eq.~\ref{eq:expand}. By the assumption $U \gg |\omega_i - U| \gtrsim |t|$, the intermediate states are dominated by those having no photons and exactly one holon and one doublon. Thus, they take the generic form $\ket{r_d; r_h; \set{\tau}} \otimes \ket{\emptyset}$, where $\ket{r_d; r_h; \{\tau\}} = (\sum_\sigma c^\dg_{r_d,\sigma} c_{r_h,\sigma}) \ket{\set{\tau}}$ is obtained from the spin state $ \ket{\{\tau\}}$ by removing an electron at $r_h$ and putting it at $r_d$, and $\ket{\emptyset}$ denotes the photon vacuum state. Henceforth we shall take the abbreviation that spins are summed implicitly within pairs of electron operators enclosed by parentheses, so that, e.g., $\cc{i}{j} = \sum_\sigma c^\dg_{i\sigma} c_{j\sigma}$.

Under this insertion, $(\E_i - H_U - H_\gamma)^{-1} = (\omega_i-U)^{-1}$ becomes a $c$-number. Moreover, recall that $H_t$ and (neglecting the photon part) $H_C$ are sums of operators of the form $\cc{i}{j}$. Once a particular term is picked for each of these sums, and given an initial spin state $\ket{\set{\sigma}}$, the resulting chain of operators automatically and uniquely determines the intermediate states (which may be 0). Thus the intermediate states can be trivially re-summed, and Eq.~\ref{eq:expand} becomes, in schematic form:

\begin{widetext}
\begin{equation} \label{eq:resummed}
\bra{\set{\sigma'}} T_\txt{R} \ket{\set{\sigma}} 
	= \nquad\sum_{i_1 j_1,i_2 j_2,\ldots}\nquad C_{i_1 j_1, i_2 j_2, \set{\sigma}}
	\bra{\set{\sigma'}} (c^\dg_{i_2} c_{j_2}) (c^\dg_{i_1} c_{j_1}) \ket{\set{\sigma}}
	+ C_{i_1 j_1, \ldots, i_3 j_3, \set{\sigma}} 
	\bra{\set{\sigma'}} (c^\dg_{i_3} c_{j_3}) (c^\dg_{i_2} c_{j_2}) (c^\dg_{i_1} c_{j_1}) \ket{\set{\sigma}} + \ldots \punct{.}
\end{equation}
\end{widetext}

The sum in Eq.~\ref{eq:resummed} is formidable. However, if $H_C$ and $H_t$ connects only between sites that are a few lattice constants away, then at low order in $t/(\omega_i-U)$, except for the choice of the initial site ($j_1$ in Eq.~\ref{eq:resummed}) the number of non-zero terms is finite and does not scale with the lattice size. Thus,  Eq.~\ref{eq:resummed} provides a systematic way of analyzing the contributions to the Raman intensity. 

The final step in the Shastry--Shraiman formulation is to convert the chain of electron operators $(c^\dg_{i_n} c_{j_n}) \cdots (c^\dg_{i_1} c_{j_1})$ into spin operators using the anti-commutation relation and the following spin identities:
\begin{equation} \label{eq:spin_identity}
\begin{aligned}
c^\dg_{\sigma} c_{\sigma'} & = \tchi_{\sigma' \sigma}
	= \frac{1}{2} \delta_{\sigma', \sigma} + \vv{S} \cdot \vvsym{\tau}_{\sigma' \sigma}
	\punct{,} \\
c_{\sigma} c^\dg_{\sigma'} & = \chi_{\sigma \sigma'}
	= \frac{1}{2} \delta_{\sigma, \sigma'} - \vv{S} \cdot \vvsym{\tau}_{\sigma \sigma'} 
	\punct{,}
\end{aligned}
\end{equation}
where $\vv{S} = c^\dg_\sigma (\vvsym{\tau}_{\sigma \sigma'}/2) c_{\sigma'}$ is the spin operator for spin-1/2 and $\vvsym{\tau}$ is the usual Pauli matrices. 

To the lowest non-vanishing order in $t/(\omega_i-U)$, the Shastry--Shraiman formulation reproduces the Fleury--London Hamiltonian,\cite{Fleury:PR:1968} i.e.:
\begin{equation} \label{eq:Fleury-London}
\begin{aligned}
\bra{f} T_\txt{R} \ket{i} & = \bra{\set{\sigma'}} H_{\txt{FL}} \ket{\set{\sigma}} + \bigO\left(\frac{t^3}{(\omega_i-U)^2}\right) \punct{,} \\
H_{\txt{FL}} & = \sum_{\vv{r},\vv{r}'} \frac{2 t_{\vv{r}\vv{r}'}^2}{U-\omega_i}  (\vv{e}_i \cdot \vvsym{\mu}) (\bar{\vv{e}}_f \cdot \vvsym{\mu}) \left(\frac{1}{4} - \vv{S}_\vv{r} \cdot \vv{S}_{\vv{r}'}\right) \punct{,}
\end{aligned} \end{equation}
where $\vvsym{\mu} = \vv{r}' - \vv{r}$ is the vector that connects lattice site $\vv{r}$ to lattice site $\vv{r}'$.

\begin{figure}
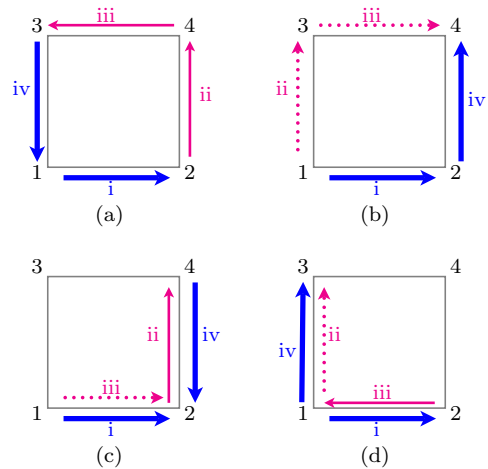

\begin{center}
\subfigure[\label{fig:sq_intro_1}]{\squareone} \qquad 
\subfigure[\label{fig:sq_intro_2}]{\squaretwo} \\
\subfigure[\label{fig:sq_intro_3}]{\squarethree} \qquad
\subfigure[\label{fig:sq_intro_4}]{\squarefour}
\caption{\label{fig:sq_intro} Pathways that contribute to the spin-chirality term in the square lattice. A thick (blue) arrow indicates the initial or the final hop, which arises from $H_C$. A thin (magenta) unbroken arrow indicates the movement of a doublon, and a thin (magenta) broken arrow indicates the movement of a holon, both arising from $H_t$.}
\end{center}
\end{figure}

For the square lattice with only nearest-neighbor hopping, \emph{Shastry and Shraiman claimed that spin-chirality term $\vv{S}_i \cdot (\vv{S}_j \times \vv{S}_k)$ appears in the $\bar{e}^{x}_{f} e^{y}_{i} - \bar{e}^{y}_{f} e^{x}_{i}$ channel at the $t^4/(\omega_i-U)^3$ order. However, our re-derivation does not confirm this result and instead concludes that the spin-chirality term vanishes to this order.} For details about our re-derivation, see Appendix~\ref{sect:A2g-derive}. While it is hard to pin down the source of this discrepancy, two possibilities are plausible. First, the pathways that contribute to the spin-chirality term includes not only those in which a doublon or holon hops through a loop [Figs.~\ref{fig:sq_intro_1} and \ref{fig:sq_intro_2}], but also those in which a holon ``chases'' a doublon or vice versa without involving a fourth site [Figs.~\ref{fig:sq_intro_3} and \ref{fig:sq_intro_4}]. These chasing pathways are non-intuitive and could be easily missed. Second, observe that for $c^\dg_{\sigma} c_{\sigma'}$, the spin indices are flipped in Eq.~\ref{eq:spin_identity} when going from electron operators to spin operators. This, together with the applications of the anti-commutation relation, can easily produce minus sign errors. 

Our conclusion that the spin-chirality term vanishes to the $t^4/(\omega_i-U)^3$ order in the one-band Hubbard model need not contradict with the experimental claim that the spin-chirality term has been observed in the cuprates,\cite{Sulewski:PRL:1991} for in the cuprates---with the holon being delocalized as Zhang-Rice singlet while the doublon being localized at the copper site---the holon and doublon hopping magnitude need not be equal. In that case, the crucial cancellation between the four pathways in Fig.~\ref{fig:sq_intro} no longer occurs. Furthermore, in the Shastry--Shraiman formalism the spin-chirality term may also be present at higher order in $t/(\omega_i-U)$ and/or when further neighbor hoppings are included. Since the ratio $t/(\omega_i-U)$ need not be small near resonance, these higher-order effects can manifest in experiments.

\begin{figure}
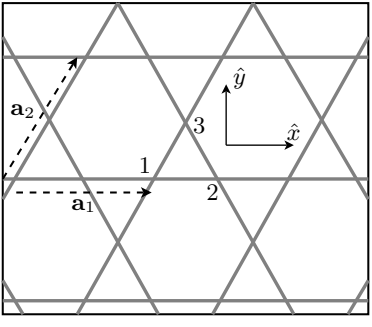

\kagomeoriginal
\caption{\label{fig:kagome_original} The kagome lattice in real space, with the unit vectors $\uv{x}$, $\uv{y}$, primitive lattice vectors $\vv{a}_1$, $\vv{a}_2$, and lattice basis $1$, $2$, $3$ defined.}
\end{figure}

Now we specialize to the kagome lattice and for simplicity assume that the hopping is between nearest neighbors only. Our convention of lattice basis, primitive lattice vectors, and axis alignment is shown in Fig.~\ref{fig:kagome_original}. \emph{In contrast to the square lattice, we found a non-vanishing spin-chirality term at the $t^4/(\omega_i-U)^3$ order.} The contrasting result between the kagome lattice and the square lattice can be traced back to the lack of four-site loop in the former.

For theoretical calculations, it is convenient to decompose the polarization dependence of the Raman intensity into the irreducible representations (irreps) of the lattice point group, since operators belonging to different irreps do not interfere with each other (note however that subtractions between various experimental setups are often required to extract the signal that corresponds to a particular channel\cite{Shastry:IJMPB:1991}). It is known\cite{Bert:JPConf:2009} that herbertsmithite belongs to the space group $R\bar{3}m$ and hence to the point group $D_{3d}$. In $D_{3d}$, the polarization tensor $\sum_{\alpha,\beta = x,y} C_{\alpha,\beta} \bar{e}^{\alpha}_{f} e^{\beta}_{i}$ in the kagome plane decomposes into two one-dimensional irreps $A_{1g}$ and $A_{2g}$, and one two-dimensional irrep $E_g$:
\begin{equation} \label{eq:irrep}
\begin{aligned}
A_{1g} & : \bar{e}^{x}_{f} e^{x}_{i} + \bar{e}^{y}_{f} e^{y}_{i} \punct{,}\\
A_{2g} & : \bar{e}^{x}_{f} e^{y}_{i} - \bar{e}^{y}_{f} e^{x}_{i} \punct{,}\\
E_{g} \left\{ \begin{array}{l} E_{g}^{(1)} \\ E_{g}^{(2)} \end{array} \right. &
	\begin{array}{l}
	: \rule{0pt}{2.5ex}\bar{e}^{x}_{f} e^{x}_{i} - \bar{e}^{y}_{f} e^{y}_{i}\\ 
	: \rule{0pt}{2.5ex}\bar{e}^{x}_{f} e^{y}_{i} + \bar{e}^{y}_{f} e^{x}_{i}
	\end{array} \punct{.}
\end{aligned}
\end{equation}

To the lowest non-vanishing order in $t/(\omega_i - U)$, the $A_{1g}$ and the $E_g$ component of the $T$-matrix are derived from the Fleury--London Hamiltonian Eq.~\ref{eq:Fleury-London}. However, the resulting expression for the $A_{1g}$ channel is the sum of a constant and a term proportional to the Heisenberg Hamiltonian and thus at zeroth order in $t/U$ does not induce any inelastic transitions. For inelastic transitions in the $A_{1g}$ channel, the leading-order contribution appears at the $t^4/(\omega_i - U)^3$ order instead. The leading-order contribution to the $A_{2g}$ channel also appears at the $t^4/(\omega_i - U)^3$ order. Explicitly, to leading order and neglecting the elastic part, the operators that corresponds to the different channels are:
\begin{widetext}
\begin{align}
O_{E_g^{(1)}} & = \frac{4t^2}{\omega_i-U} \sum_{R} \bigg(
	\frac{1}{4} \vv{S}_{R,3} \cdot 
	\left( \vv{S}_{R,1} + \vv{S}_{R+\atwo,1} + 
	\vv{S}_{R,2} + \vv{S}_{R+\atwo-\aone,2}	\right) 
	-	\frac{1}{2} \vv{S}_{R,2} \cdot \left(
	\vv{S}_{R,1} + \vv{S}_{R+\aone,1} \right)
	\bigg) \punct{,} \label{eq:Eg1} \\
O_{E_g^{(2)}} & = \frac{4t^2}{\omega_i-U} \sum_{R} \frac{\sqrt{3}}{4} \Big(
	\vv{S}_{R,3} \cdot (\vv{S}_{R,1} + \vv{S}_{R+\atwo,1}) 
	- \vv{S}_{R,3} \cdot (\vv{S}_{R,2} + \vv{S}_{R+\atwo-\aone,2})
	\Big) \punct{,} \label{eq:Eg2} \\
O_{A_{1g}} & = \frac{-t^4}{(\omega_i-U)^3} \sum_{R} 2 \Big(
	\vv{S}_{R,1} \cdot (\vv{S}_{R+\aone,1} + \vv{S}_{R+\atwo,1}) 
	+ \vv{S}_{R,2} \cdot (\vv{S}_{R-\aone,2} + \vv{S}_{R+\atwo-\aone,2}) 
	+ \vv{S}_{R,3} \cdot (\vv{S}_{R-\atwo,3} + \vv{S}_{R-\atwo+\aone,3}) 
	\Big) \notag \\
	& \qquad + \Big(
	\vv{S}_{R,1} \cdot (\vv{S}_{R+\atwo-\aone,2} + \vv{S}_{R+\aone-\atwo,3}) 
	+ \vv{S}_{R,2} \cdot (\vv{S}_{R-\atwo,3} + \vv{S}_{R+\atwo,1}) 
	+ \vv{S}_{R,3} \cdot (\vv{S}_{R+\aone,1} + \vv{S}_{R-\aone,2}) 
		\Big) \punct{,} \label{eq:A1g} \\
O_{A_{2g}} & = \frac{2\sqrt{3}it^4}{(\omega_i-U)^3} \sum_{R} \Big(
	3 \SSS{R,1}{R,2}{R,3}  + 3 \SSS{R,1}{R-\aone,2}{R-\atwo,3}
	+ \SSS{R,1}{R,3}{R+\atwo-\aone,2} + \SSS{R+\atwo,1}{R,3}{R,2} \notag \\
	& \rule{0.05\textwidth}{0pt} + \SSS{R,3}{R,2}{R+\aone,1} + 
	\SSS{R-\atwo+\aone,3}{R,2}{R,1} + \SSS{R,2}{R,1}{R-\atwo,3} 
	+ \SSS{R-\aone,2}{R,1}{R,3}
	\Big) \notag \\
& = \frac{2\sqrt{3}it^4}{(\omega_i-U)^3} \sum_R 
	\Big( 3 \uptri + 3 \dntri + \wwtov + \wtovv + \vvtox + \vtoxx + \xxtow + \xtoww \Big) \punct{,} \label{eq:A2g} 
\end{align}
\end{widetext}
\noindent where $\SSS{i}{j}{k}$ denotes $\vv{S}_i \cdot (\vv{S}_j \times \vv{S}_k)$. Further discussion on the derivation of these expressions can be found in the Appendices.

\section{U(1) Dirac spin-liquid state} \label{sect:spin_liquid}

At half filling ($\avg{\sum_\sigma n_{i\sigma}} = 1$), on the near-degenerate ground-state manifold and to the leading order in $t/U$, the Hubbard model is reduced to the Heisenberg model:
\begin{equation} \label{eq:Heisenberg}
H_{\textrm{Hsb}} = \frac{1}{2} \sum_{ij} J_{ij} \left(\SdotS{i}{j} - \frac{1}{4} \right) \punct{,}
\end{equation}
where $J_{ij} = 4 t^2_{ij}/U$. As in the previous section, we shall specialize to the case where the hopping is between nearest neighbor only, and hence $J_{ij} = 0$ unless $i$, $j$ are nearest neighbor.

In a spin-liquid state $\avg{\vv{S}_i} = 0$. Thus, new variables are needed to describe the order of the system. A common choice is to introduce fermionic spinon operators $f_{i\sigma}$, $f^\dg_{i\sigma}$ to decompose the spin operator $\vv{S}_i$ as follows:
\begin{equation} \label{eq:S-to-f}
\vv{S}_i = \frac{1}{2} \sum_{\alpha,\beta} f^\dg_{i\alpha} \vvsym{\tau}_{\alpha,\beta} f_{i\beta} \punct{.}
\end{equation} 
The occupation constraint $\sum_\sigma n_{i\sigma} = 1$ resulting from the $U \rightarrow \infty$ limit then becomes the constraint $\sum_\sigma f^\dg_{i\sigma} f_{i\sigma} = 1$ on the spinons.\cite{Baskaran:SSC:1987, Kotliar:PRB:1988}

Under this transformation, the spin-spin interaction $\SdotS{i}{j}$ is mapped to a four-fermion interaction, which can be decoupled using a Hubbard--Stratonovich transformation. This yields the partition function $Z = \int Df Df^\dagger D\lambda D\chi \exp\left(-\int_0^\beta d\tau L\right)$, with the Lagrangian $L$ given by:
\begin{equation} \label{eq:Lagrangian}
\begin{aligned}
L & = J \sum_{\avg{ij}} \left( |\chi_{ij}|^2 - 
	\sum_\sigma (\chi_{ij}^* f^\dg_{i\sigma} f_{j\sigma} + c.c.) \right) \\
	& \quad + \sum_{i\sigma} f^\dagger_{i\sigma} (\partial_\tau - i \lambda_i) f_{i\sigma} \punct{.}
\end{aligned}
\end{equation}

\begin{figure}
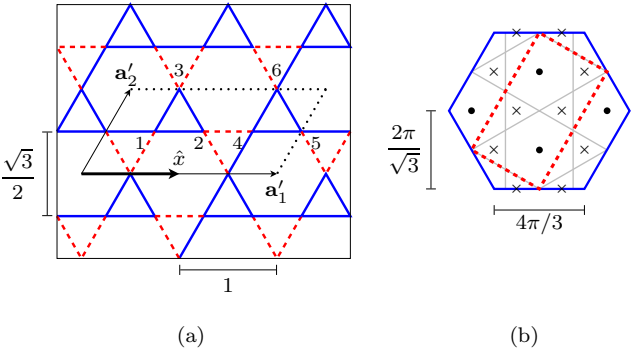

	\subfigure[\label{fig:kagome_doubled}]{\kagomedouble} \,
	\subfigure[\label{fig:kagome_kspace}]{\kagomebrillouin} 
	\caption{\label{fig:lattice} (a) The kagome lattice with the DSL ansatz. The (red) dashed lines correspond to bonds with effective hopping $\tilde{t} = -\chi J$ while the (blue) unbroken lines correspond to bonds with effective hopping $\tilde{t} = \chi J$. $\vv{a}'_1$ and $\vv{a}'_2$ are the primitive vectors of the doubled unit cell. (b) The original Brillouin zone (bounded by unbroken lines) and the reduced Brillouin zone (bounded by broken lines) of the DSL ansatz. The dots indicate locations of the Dirac nodes at half-filling, the crosses indicate locations of the Dirac nodes crossing the second and the third band, and the thin (gray) lines indicate saddle regions at which the band energies are the same as that at $\vv{k} = 0$.}
\end{figure}

Mean-field ansatzs can be specified by treating $\chi_{ij}$ as an order parameter. Since a spin-liquid state is invariant under translation and rotation, $|\chi_{ij}|$ must be independent of $i$, $j$. Letting  $\chi_{ij} = \chi e^{-i a_{ij}}$ and rewriting $\lambda_i = a_0^i$, the Lagrangian Eq.~\ref{eq:Lagrangian} yields the following mean-field Hamiltonian:
\begin{equation} \label{eq:H_MF}
H_{\textrm{MF}}\! = \! 
\sum_{i\sigma} \opdag{f}_{i\sigma}(i a_0^i - \mu_F) \op{f}_{i\sigma}
	\!-\! \chi J \!\!\sum_{\avg{ij},\sigma}\! 
	(e^{ia_{ij}}\opdag{f}_{i\sigma}\op{f}_{j\sigma} \!+\! \txt{h.c.}) \punct{.} 
\end{equation}

Observe that an internal gauge field $a^\mu$ emerges naturally from this formulation. Its space components $a_{ij}$ arise from the phases of $\chi_{ij}$, while its time component $a_0$ arises from enforcing the occupation constraint.

By gauge invariance, a mean-field ansatz for $a^\mu$ is uniquely specified by the amount of fluxes through the triangles and the hexagons of the kagome lattice. In particular, the DSL state is characterized by zero flux through the triangles and $\pi$ flux through the hexagons.\cite{Ran:PRL:2007,Hermele:PRB:2008,Hastings:PRB:2001} By picking an appropriate gauge, the DSL state can be described by a tight-binding Hamiltonian with effective hopping $\tilde{t} = \pm \chi J$ on each bond. For the precise pattern of the $\pm$ signs see Fig.~\ref{fig:kagome_doubled}. Note that the unit cell for this effective tight-binding Hamiltonian is necessarily doubled as the flux enclosed by the original unit cell is $\pi$.

\begin{figure}
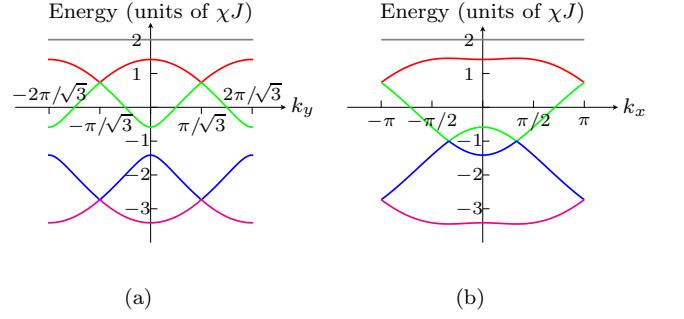

	\subfigure[\label{fig:kx=0}]{\kagomeyplot} \,
	\subfigure[\label{fig:ky=0}]{\kagomexplot}
	\caption{\label{fig:Brillouin} The band structure of the kagome lattice with the DSL ansatz plotted along  (a) $k_x = 0$ and (b) $k_y = 0$. Note that the top band is twofold degenerate.}
\end{figure}

It is easy to check that, in units where $\chi J = 1$, this effective tight-binding Hamiltonian produces the following bands:
\begin{align}
E_{\textrm{top}} & = 2 \qquad \textrm{(doubly degenerate)} \punct{,} \label{eq:topband}\\
E_{\pm,\mp} & = -1 \pm \sqrt{ 3 \mp \sqrt{2} 
	\sqrt{3-\cos 2 k_x + 2 \cos k_x \cos\sqrt{3} k_y} } \punct{.} \label{eq:Diracbands}
\end{align}
At any $\vv{k}$-point, $E_{-,+} \leq E_{-,-} \leq E_{+,-} \leq E_{+,+} < E_{\textrm{top}}$. For plots of this band structure, see Fig.~\ref{fig:Brillouin}.

At low energy, the spinon spectrum is well-described by four (two spins times two $\vv{k}$-points) Dirac nodes, located at momentum $\pm \vv{Q} = \pm \pi/\sqrt{3} \, \uv{y}$ [c.f.\@ Figs.~\ref{fig:kagome_kspace} and \ref{fig:kx=0}]. More specifically, at low-energy we may replace the mean-field Hamiltonian Eq.~\ref{eq:H_MF} by the Dirac Hamiltonian:
\begin{equation} \label{eq:H_Dirac}
H_{\textrm{Dirac}} = \nu_F \sum_{\sigma,\alpha,\vv{q}} \psi^\dg_{\sigma,\alpha,\vv{q}} (q_x \tau_x + q_y \tau_y) \psi_{\sigma,\alpha,\vv{q}} \punct{,}
\end{equation}
where $\sigma = \up, \dn$ index spins, $\alpha = \pm$ index the location of the Dirac node, and $\vv{q}$ denotes the momentum as measured from the Dirac node. The relation between the two-component fermionic operators $\psi, \psi^\dg$ and the spinon operators $f$,$f^\dg$ defined on the lattice sites can be found in Ref.~\onlinecite{Hermele:PRB:2008}.

As will be explained in details below, the Dirac structure of this low-energy Hamiltonian bears important consequences for the Raman intensity at low energy. In particular, \emph{the power-law behavior of the Raman intensity at low energy is largely resulted from the strong phase space restriction of the Dirac node structure.}

Under the DSL ansatz, we may take $\ket{i} = \ket{i^{\Hbb}} \otimes  \ket{\vv{k}_i,\vv{e}_i}$ and $\ket{f} = \ket{f^{\Hbb}} \otimes  \ket{\vv{k}_f,\vv{e}_f}$ in the transition rate Eq.~\ref{eq:rate}, where $\ket{i^{\Hbb}}$ and $\ket{f^{\Hbb}}$ are states obtained from filling the spinon bands. In particular, at zero temperature (which will be assumed henceforth), $\ket{i^\Hbb}$ is simply a state with spinons filled up to the Dirac nodes at the top of $E_{+,-}$, and $\ket{f^\Hbb}$ are states with a few spinon-antispinon pairs excited from $\ket{i^\Hbb}$.

Moreover, the spin operators that appear in Eqs.~\ref{eq:Eg1}--\ref{eq:A2g} can be converted to spinon operators using Eq.~\ref{eq:S-to-f}. Explicitly,
\begin{align}
\SdotS{i}{j} & = \frac{1}{4} - \frac{1}{2} 
	f^\dg_{i\sigma} f_{j\sigma} f^\dg_{j\sigma'} f_{i\sigma'} \punct{,} \label{eq:SS} \\
\vv{S}_{i} \cdot (\vv{S}_j \times \vv{S}_k) & = \frac{i}{4} \left(
	\ff{i}{j} \ff{j}{k} \ff{k}{i} - \txt{h.c.} \right) \punct{,} \label{eq:SSS}
\end{align}
which allows the matrix elements between $\ket{i^{\Hbb}}$ and $\ket{f^{\Hbb}}$ to be calculated at mean field using Wick's theorem.

At zero temperature, to obtain the overall Raman intensity $I_{\alpha}$ for channel $\alpha$ at a given Raman shift $\Delta\omega = \omega_i - \omega_f$, all final states that satisfies the energy constraint $\E^{\Hbb}_f - \E^{\Hbb}_i = \Delta\omega$ must be summed over. Strictly enforcing this constraint is difficult in a numerical computation. Instead, we sample the final states $\ket{f^{\Hbb}}$ without imposing the energy constraint, but sort them into bins of energy. The overall intensity is then obtained by summing all states whose energy fall within the same bin. We perform the summation numerically using a simple Monte Carlo sampling of the momenta and band indices of the spinon and antispinon excitations in $\ket{f^{\Hbb}}$.

Some calculations are also performed in a slightly different way, by first converting the summed squared amplitude into a correlation function:
\begin{equation} \label{eq:correlation} \begin{aligned}
I_{\alpha}(\Delta\omega) & = \sum_f \Big|\bra{f^\Hbb} O_\alpha \ket{i^\Hbb}\Big|^2 \delta(\E^\Hbb_f - \E^\Hbb_i-\Delta\omega) \\
& = \int dt \: e^{i\Delta\omega t} \bra{i^\Hbb} O_\alpha(t) O_\alpha(0) \ket{i^\Hbb}
\punct{.}
\end{aligned}\end{equation}
To calculate this correlation function, the spinon operators that appears in $O_\alpha$ (c.f.\@ Eqs.~\ref{eq:Eg1}--\ref{eq:A2g}) are converted from real space to momentum space, which introduces sums over momenta and band indices. Such sums are again computed by simple Monte Carlo samplings in the manner explained above.

To simplify notations, in the following we shall abuse notation and write $\ket{i}$ in place of $\ket{i^\Hbb}$ and similarly write $\ket{f}$ for $\ket{f^\Hbb}$ whenever the context is clear.

\section{\texorpdfstring{$E_g$}{Eg} channel} \label{sect:Eg}

\begin{figure}[h]
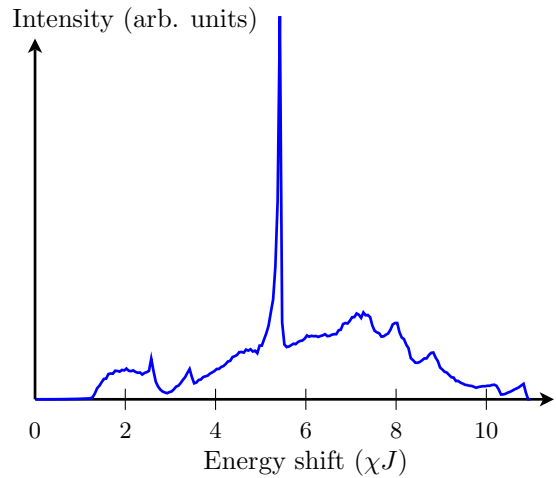

\begin{center}
\Egcorrplot
\caption{\label{fig:Eg_corr}Raman intensity in the $E_g$ channel, computed using the correlation Eq.~\ref{eq:correlation}, with a bin size of $0.05 \chi J$ in energy.}%
\end{center}
\end{figure}

First consider the Raman intensity in the $E_g$ channel, $I_{E_g} \equiv I_{E_g^{(1)}}+I_{E_g^{(2)}}$. By computing the correlation function Eq.~\ref{eq:correlation} (with $\alpha$ = $E_g^{(1)}$ and $E_g^{(2)}$), we obtain the Raman intensity profile as shown in Fig.~\ref{fig:Eg_corr}.

From Fig.~\ref{fig:Eg_corr}, it can be seen that the Raman response take the form of a broad continuum ranging from approximately $1.5 \chi J$ to approximately $11 \chi J$, with occasional sharp spikes. The existence of a continuum is not a surprise, since the operators $O_{E_g^{(1)}}$ and $O_{E_g^{(2)}}$ in Eqs.~\ref{eq:Eg1}--\ref{eq:Eg2} corresponds to two-spinon-two-antispinon operators in the DSL ansatz, and hence the final states $\ket{f}$ are spinon-antispinon pairs and thus there is a continuum of phase space for excitations. The cutoff near $11 \chi J$ is also natural, since the total band-width in the DSL ansatz is approximately 6 $\chi J$, and hence with two spinons and two antispinons the excitation energy is at most 12 $\chi J$. The sharp peaks and the low-energy suppression, however, require further investigations.

To gain more insight into how the various features in Fig.~\ref{fig:Eg_corr} comes about, it is useful to consider the sum over final states explicitly. Given the form of $O_{E_g^{(1)}}$ and $O_{E_g^{(2)}}$ in Eqs.~\ref{eq:Eg1}--\ref{eq:Eg2}, there are two types of final states: the first consists of two spinon-antispinon pairs excited from the ground state $\ket{i}$ while the second consists of \emph{one} spinon-antispinon pair excited from $\ket{i}$. Schematically, these take the form:
\begin{align}
\ket{f_{\txt{2pairs}}} & = f^\dg_{k_1,n_1,\sigma} f_{k_2,n_2,\sigma} f^\dg_{k_3,n_3,\sigma'} f_{k_4,n_4,\sigma'} \ket{i} \punct{,} \label{eq:2pairs} \\
\ket{f_{\txt{1pair}}} & = f^\dg_{k_1,n_1,\sigma} f_{k_2,n_2,\sigma} \ket{i} \punct{.} \label{eq:1pair}
\end{align}
In both cases the momentum conservation $\sum_{i\txt{ odd}} \vv{k}_i= \sum_{i\txt{ even}} \vv{k}_i$ holds, and that the band index $n_i$ denotes one of the top three (empty) bands when $i$ is odd and one of the bottom three (occupied) bands when $i$ is even.

For the one-pair final state to have a non-zero matrix element with the initial state, a pair of spinon-antispinon operator must self-contract within $O_{E_g^{(1)}}$ and $O_{E_g^{(2)}}$, e.g.,
\begin{equation} \label{eq:contract}
\begin{aligned}
f^\dg_{i\sigma} f_{j\sigma} f^\dg_{j\sigma'} f_{i\sigma'}
& \mapsto \avg{f^\dg_{i\sigma} f_{j\sigma}} f^\dg_{j\sigma'} f_{i\sigma'}
	+ f^\dg_{i\sigma} f_{j\sigma} \avg{f^\dg_{j\sigma'} f_{i\sigma'}} \\
& \quad - \avg{f^\dg_{i\sigma} f_{i\sigma'}} f^\dg_{j\sigma'} f_{j\sigma}
	- \avg{f^\dg_{j\sigma'} f_{j\sigma}} f^\dg_{i\sigma} f_{i\sigma'} \\
& \nquad \nquad = \chi_{ij} f^\dg_{j\sigma} f_{i\sigma} + \chi_{ji} f^\dg_{i\sigma} f_{j\sigma}
	- \frac{1}{2} ( f^\dg_{i\sigma} f_{i\sigma} + f^\dg_{j\sigma} f_{j\sigma}) \punct{,}
\end{aligned}
\end{equation}
where the spin indices $\sigma$, $\sigma'$ are summed in the above. The mean-field parameter $\chi_{ij}$ here is the same as the one introduced in Sec.~\ref{sect:spin_liquid}.

\begin{figure}
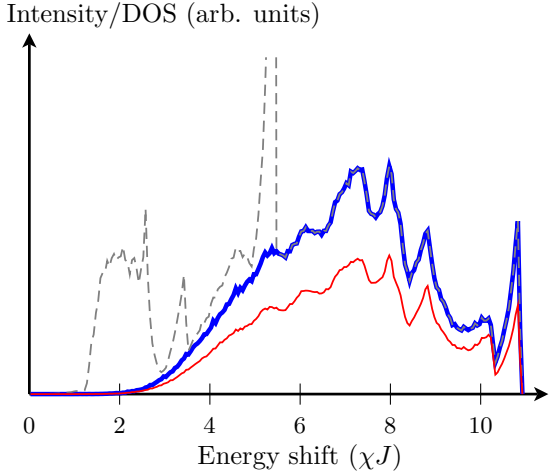

\begin{center}
\twopairplot
\caption{\label{fig:2pairs}Plots of density of states (thin red curve) and $E_g$ Raman intensity (thick blue curve) contributed by final states having two spinon-antispinon pairs, together with the overall $E_g$ Raman intensity (broken gray curve), all computed with a bin size of $0.05 \chi J$ in energy.  Note that the relative scale between the intensity plots and the DOS plot is arbitrary and is set here such that both curves are visible.}
\end{center}
\end{figure}

\begin{figure}
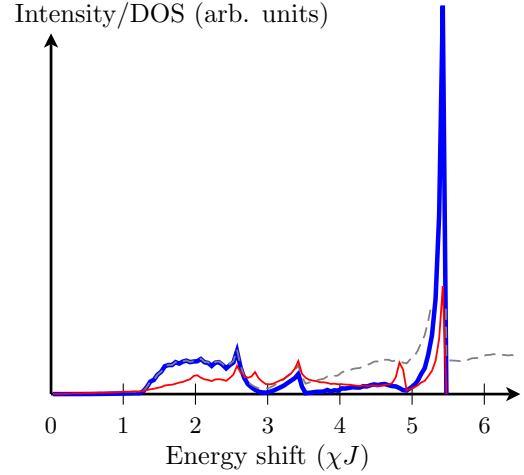

\begin{center}
\onepairplot
\caption{\label{fig:1pair}Plots of density of states (thin red curve) and $E_g$ Raman intensity (thick blue curve) contributed by final states having one spinon-antispinon pair, together with the overall $E_g$ Raman intensity (broken gray curve), all computed with a bin size of $0.05 \chi J$ in energy. Note that the relative scale between the intensity plots and the DOS plot is arbitrary.}
\end{center}
\end{figure}

However, one should be mindful that while the constraint $\sum_\sigma f^\dg_{j\sigma} f_{j\sigma} = 1$ is strictly enforced in the exact spin-liquid state, it is enforced only on average in the mean-field representation. Therefore, \emph{while operators of the form $f^\dg_{j\sigma} f_{j\sigma}$ should induce no transition between $\ket{i}$ and $\ket{f}$, the matrix element $\bra{f} f^\dg_{j\sigma} f_{j\sigma} \ket{i}$ may be non-zero at mean-field.} To avoid such problem, we shall throw away the $f^\dg_{j\sigma} f_{j\sigma}$ terms in Eq.~\ref{eq:contract} by hand.

With this precaution, we recalculated $I_{E_g}$ by numerically sampling the final states, and we do so for the one-pair and two-pair contributions separately. The results are shown on Figs.~\ref{fig:2pairs} and \ref{fig:1pair}. We have verified that the sum of Raman intensities from the two figures (with the relative ratio determined by the details of the numerical calculations) produces an overall intensity profile that matches Fig.~\ref{fig:Eg_corr}.

The density of states (DOS) for two-pair and one-pair excitations having zero total momenta have been plotted alongside with the respective Raman intensities in Figs.~\ref{fig:2pairs} and \ref{fig:1pair}. From the figures, it can be seen that the DOS matches the Raman intensity profile very well for two-pair excitations, and less so (but still reasonably well) for one-pair excitations. This can be understood by rewriting the first line of Eq.~\ref{eq:correlation} as $I_\alpha(\Delta\omega) = \overline{\big| \bra{f} O_\alpha \ket{i} \big|^2} \DOS(\Delta\omega)$, in which $\DOS(\E)$ denotes the density of state at energy $\E$ and $\overline{\big| \bra{f} O_\alpha \ket{i} \big|^2}$ denotes the average matrix element squared at the same energy. From this, the DOS is expected to match the Raman intensity well as long as the average matrix element does not change drastically with energy---an assumption more valid for two-pair states as opposed to one-pair ones, because of the larger phase space available in the former case.

Moreover, the sharp peak appearing near $5.5 \chi J$ in Fig.~\ref{fig:Eg_corr} can now be attributed to one-pair excitations, which can in turn be attributed to a peak in the one-pair DOS. From the band structure (Fig.~\ref{fig:Brillouin}), it can be checked that \emph{$\Delta\omega = 5.41 \chi J$ corresponds to the energy difference between the top flat band and the saddle point $\vv{k} = \vv{0}$ at the bottom band.} The enhanced phase space near this energy is thus the likely cause of this peak-like feature.

\begin{figure}
\includegraphics[]{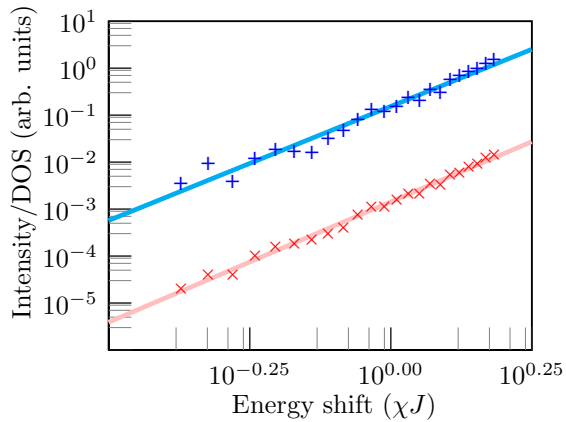}
\caption{\label{fig:2pairslnln}Log-log plot of the DOS (red $\times$ symbols) and Raman intensity in the $E_g$ channel (blue $+$ symbols) for two-pair excitations. Simple linear fits (straight lines in pink and cyan) give a slope of $5.1$ for the DOS data and $4.9$ for the Raman intensity data.}
\end{figure}

\begin{figure}
\includegraphics[]{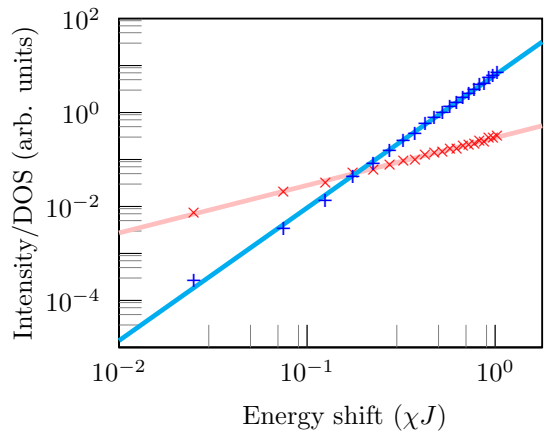}
\caption{\label{fig:1pairlnln}Log-log plot of the DOS (red $\times$ symbols) and Raman intensity in the $E_g$ channel (blue $+$ symbols) for one-pair excitations. Simple linear fits (straight lines in pink and cyan) give a slope of $1.0$ for the DOS data and $2.8$ for the Raman intensity data.}
\end{figure}

Next, we consider the low-energy ($\Delta\omega < 2 \chi J$) part the of Raman intensity more carefully, checking if the intensity profile is an exponential or a power law, and determining the exponent if the latter case holds. To enhance the quality of the data, and since only the low-energy behavior concerns us, we resample the final states, restricting the antispinon to the highest occupied band $E_{+,-}$ and the spinon to the lowest unoccupied band $E_{+,+}$ (despite this, the two-pair data is consistently non-zero only above $\Delta\omega = 0.4 \chi J$). The resulting data are shown in the log-log plots in Figs.~\ref{fig:2pairslnln} and \ref{fig:1pairlnln}. Since the smaller exponent dominates as $\Delta\omega \rightarrow 0$, the \emph{overall Raman intensity in the $E_g$ channel scales roughly as $I_{E_g} \propto (\Delta\omega)^3$ at low energy.}

From the plots, it is clear that the DOS and the Raman intensity both follow a power law, with a higher exponent for the two-pair excitations as compared to the one-pair ones. \emph{Analytically, it is easy to check that $\DOS_{\txt{1pair}} \propto \Delta\omega$ for the Dirac Hamiltonian Eq.~\ref{eq:H_Dirac}}, since the integral involves four components of momenta subjected to three constraints (energy and momentum conservation), and that $|\vv{q}|$ scales as energy. \emph{Similarly, it is easy to check that $\DOS_{\txt{2pair}} \propto (\Delta\omega)^5$ in the Dirac Hamiltonian.} Furthermore, since the eigenstates of the Dirac Hamiltonian depend only on $\theta = \tan^{-1}(q_y/q_x)$ but not on $|\vv{q}|$, \emph{the average matrix element squared $\overline{\big| \bra{f} O_\alpha \ket{i} \big|^2}$ must be constant in energy.} However, to the $t^2/(\omega_i-U)$ order in the $E_g$ channel (Eqs.~\ref{eq:Eg1}--\ref{eq:Eg2}), the matrix element turns out to be exactly zero for all one-pair excitations in the Dirac Hamiltonian. Hence, at low energy, we expect $I_{\txt{2pair}} \propto \DOS_{\txt{2pair}} \propto (\Delta\omega)^5$ and $I_{\txt{1pair}} \propto (\Delta\omega)^\alpha$ with $\alpha > 1$, consistent with the numerical results in Figs.~\ref{fig:2pairslnln} and \ref{fig:1pairlnln}.

Since $\DOS_{\txt{1pair}} \propto \Delta\omega$, it leaves open the possibility that $I_{E_g} \propto \Delta\omega$ at higher order in $t/(\omega_i-U)$. While we are not able to rule out this possibility, we do find that \emph{the vanishing of matrix elements in the Dirac Hamiltonian for all one-pair excitations persist to the $t^4/(\omega_i-U)^3$ order} (the derivation of $O_{E_g}$ to this order can be found in Appendix~\ref{sect:Eg-derive}).\footnote{To perform the analytic calculations, the expectation \protect{$\avg{f^\dagger_i f_j}$} for next- and next-next-nearest neighbors are needed. From the DSL ansatz these must be real, equal in magnitude, but vary in signs. We determine the signs using numerical calculations as a shortcut. It turns out that the cancellation holds separately for next-nearest neighbor terms and next-next-nearest ones. Thus the determination of signs is adequate for reaching our conclusion.} Hence, even if $I_{E_g} \propto \Delta\omega$ at some higher order, its effect will not be prominent unless the system is sufficiently near resonance.

In the data presented above we have summed the contributions arising from $O_{E_g^{(1)}}$ and $O_{E_g^{(2)}}$. By computing the two contributions separately, it can be checked that each contribute equally. In fact, we have checked that \emph{the intensity profiles are essentially identical upon an arbitrary rotation in the kagome plane for the one-pair and two-pair excitations separately}. In other words, we found that the quantity $I_{E_g}(\Delta\omega,\theta)$, defined by:
\begin{equation}\label{eq:rotation}
\begin{aligned}
I_{E_g}(\Delta\omega,\theta) & = \sum_{f} |\bra{f}O_{E_g^{(1)}} \cos\theta + O_{E_g^{(2)}} \sin\theta \ket{i}|^2 \\
& \qquad \times \delta(\E^\Hbb_f - \E^\Hbb_i-\Delta\omega) \punct{,}
\end{aligned}
\end{equation}
is independent of $\theta$, and remain so even if the sum is restricted to one-pair or two-pair states. Our numerical results are thus consistent with the analytical arguments given by Cepas \etal\cite{Cepas:PRB:2008}

\section{\texorpdfstring{$A_{1g}$}{A1g} channel} \label{sect:A1g}

\begin{figure}[ht]
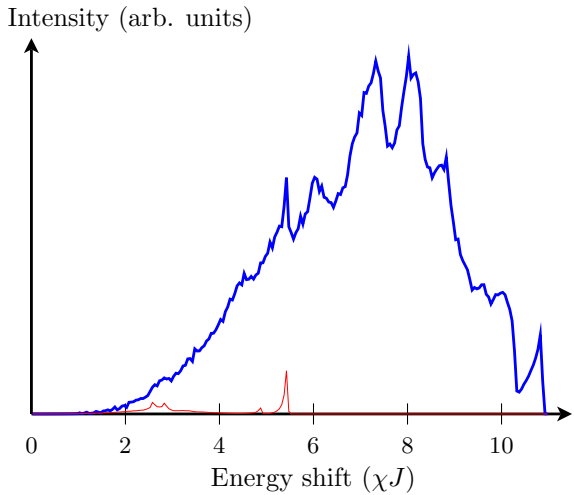

\Aonegsumplot
\caption{\label{fig:A1g}Plots of the overall Raman intensity (thick and blue) and its contribution by one-pair states (thin and red) in the $A_{1g}$ channel, both with the same vertical scale and computed with a bin size of $0.05 \chi J$ in energy.}
\end{figure}

Using $O_{A_{1g}}$ in Eg.~\ref{eq:A1g} in place of $O_{E_{g}^{(1)}}$ and $O_{E_{g}^{(2)}}$, we repeat the calculation of the Raman intensity profile for the $A_{1g}$ channel. The results are shown in Fig.~\ref{fig:A1g}. From the figure, we see that the Raman intensity profile of the $A_{1g}$ channel also has a broad continuum up to a cutoff near $11 \chi J$. However, the sharp peak near $5.5\chi J$ that appears in the $E_g$ channel is markedly missing. 

Decomposing the Raman intensity into one-pair and two-pair contributions as in Sec.~\ref{sect:Eg}, it can be seen that \emph{the overall Raman intensity profile in the $A_{1g}$ channel is dominated by the two-pair states.} Since the sharp peak near $5.5\chi J$ is originated from the one-pair contribution, this explains the absence of sharp peak in the $A_{1g}$ channel.

\begin{figure}
\includegraphics[]{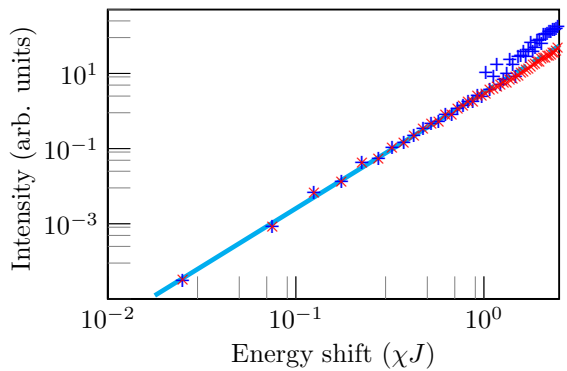}
\caption{\label{fig:A1glnln}Log-log plot of the Raman intensity (blue $+$ symbols) and its contribution by one-pair states (red $\times$ symbols) in the $A_{1g}$ channel. Simple linear fit (straight line in cyan) for the one-pair data up to $\Delta\omega = \chi J$ gives a slope of $3.1$.}
\end{figure}

However, at low energy ($\Delta\omega \lesssim 1.5 \chi J$) the Raman intensity profile is still dominated by the one-pair contribution. And by plotting the Raman intensity profile in a log-log scale (Fig.~\ref{fig:A1glnln}), we see that the low-energy behavior is characterized by a power law with exponent $\alpha \approx 3$, similar to the value obtained in the $E_g$ channel. Again, it can be checked on analytical ground\footnotemark[\value{footnote}] that the matrix element vanishes for all one-pair state in the Dirac Hamiltonian, consistent with the numerical results.

\section{\texorpdfstring{$A_{2g}$}{A2g} channel} \label{sect:A2g}

\begin{figure}[ht]
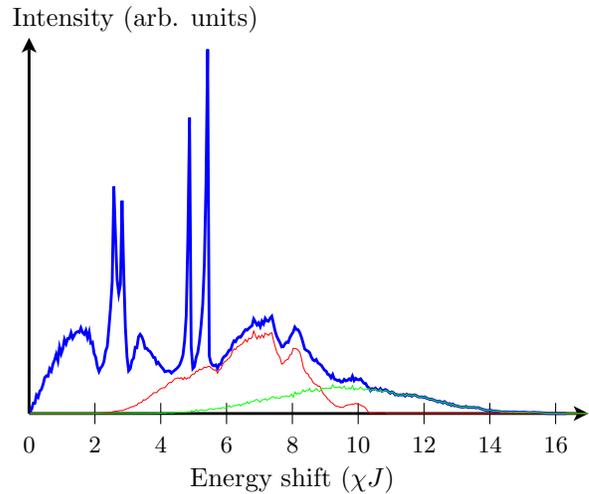

\Atwogsumplot
\caption{\label{fig:A2g}Raman intensity in the $A_{2g}$ channel arising from spinon-antispinon pairs (thick and blue), and its contribution from two-pair (thin and red) and three-pair (thin and green) states. All computed with a bin size of $0.05\chi J$ in energy.}
\end{figure}

For the $A_{2g}$ channel, to the leading order in $t/(\omega_i-U)$, the Raman process receives contributions from excited states having one, two, and three pairs of spinon-antispinon. The overall Raman intensity coming from these spinon-antispinon pairs are plotted in Fig.~\ref{fig:A2g}. It can be seen that the continuum also appears in this channel, now ranging from $0\chi J$ up to approximately $16 \chi J$, which corresponds to approximately three times the total spinon bandwidth. Moreover, sharp peaks not unlike the one in the $E_g$ channel are observed at various energies. Again, it can be checked that these sharp peaks can be attributed to the various features of the DSL bands, particularly to the flat-band-to-saddle and saddle-to-saddle transitions.

Note that \emph{the low-energy Raman transition is much more prominent in the $A_{2g}$ channel} than that in the $E_g$ and $A_{1g}$ channels. In particular, the Raman intensity in this channel has a broad peak near $\Delta\omega = 1.5 \chi J$, below which it is visibly linear. From Fig.~\ref{fig:A2g}, the prominence of low-energy Raman transition trace back to the relatively large average matrix elements in the one-pair transitions. Moreover, the linear behavior suggests that $\overline{\big| \bra{f} O_\alpha \ket{i} \big|^2}$ no longer vanishes in the Dirac Hamiltonian, so that $I_{\txt{1pair}} \propto \DOS_{\txt{1pair}} \propto \Delta\omega$ at low energy.

The forgoing discussion neglected an important contribution to the $A_{2g}$ Raman intensity. Recall that in computing the one-pair contribution, a mean-field factorization is performed (Eq.~\ref{eq:contract}), in which the Hubbard-Stratonovich variable $\chi_{ij}$ is treated as a constant. However, because of the emergent gauge structure in the U(1) spin-liquid theory Eq.~\ref{eq:H_MF}, the dynamics of the phase of $\chi_{ij}$ cannot really be neglected. Fortunately, since the projection of the $T$-matrix onto the $E_g$ and $A_{1g}$ channel Eqs.~\ref{eq:Eg1}--\ref{eq:A1g} involves no non-trivial closed path, the results presented above should still be \emph{qualitatively} correctly, even though quantitative corrections to the detailed predictions (such as the exponents of the power laws at low-energy) may be present. 

The situation for the $A_{2g}$ channel is different, as can be seen by considering the full contraction of the spin-chirality term $\vv{S}_1 \cdot ( \vv{S}_2 \times \vv{S}_3)$ into $\chi_{ij}$:
\begin{equation} \label{eq:gaugeflux}
\begin{aligned}
i \vv{S}_1 & \cdot ( \vv{S}_2 \times \vv{S}_3) 
	= \frac{1}{4} \left(	\ff{1}{3} \ff{3}{2} \ff{2}{1} - \txt{h.c.} \right) \\
	& \mapsto \frac{1}{4} \left( \chi^{(0)}_{13} \, e^{i a_{13}} 
	\chi^{(0)}_{32} \, e^{i a_{32}} \chi^{(0)}_{21} \, e^{i a_{21}} - \txt{h.c.}
	\right) + \ldots \\
	& = (C \, e^{i (a_{13} + a_{32} + a_{21})} -\txt{h.c.}) + \ldots \punct{,}
\end{aligned}
\end{equation}
where $\chi_{ij}^{(0)}$ denotes the part of $\chi_{ij}$ that can be treated as constant, and $C = \chi^{(0)}_{13} \chi^{(0)}_{32} \chi^{(0)}_{21} /4$. For convenience we shall denote $(C \, e^{i (a_{ik} + a_{kj} + a_{ji})} - \txt{h.c.})$ as $Q_{ijk}$ henceforth.

Note that $e^{i (a_{13} + a_{32} + a_{21})}$ is the emergent gauge flux enclosed by the loop $1 \rightarrow 3 \rightarrow 2 \rightarrow 1$. \emph{Since the emergent gauge field is fluctuating, this gauge flux can lead to an excitation of the kagome system.} Physically, the excitations generated by $Q$ can be thought of as a \emph{collective} excitation in the system, which exists on top of the individual spinon-antispinon excitations. This is analogous to the situation in an ordinary Fermi liquid, where plasmon mode exists on top of the electron-hole continuum.

Since the fluctuation of $a_{ij}$ can be considered as a collective excitation in the system, as an approximation we may consider the final states $\ket{f_\txt{gauge}}$ connected to ground state by $Q$ to be lying in a separate sector than the spinon-antispinon pairs previously considered. Then, the total Raman intensity in the $A_{2g}$ channel can be obtained by adding the contributions by these collective states to the contributions by the spinon-antispinon pair states.

The contribution to the Raman intensity by $\ket{f_\txt{gauge}}$ can be computed from the $QQ$ correlator. To do so, we start with Eq.~\ref{eq:gaugeflux}, take the continuum limit, and Taylor expand the exponential. Then, $a_{13} + a_{32} + a_{21} \mapsto \oint_{\wp_{132}} \vv{a} \cdot d\vv{x} = \iint_{\Omega_{132}} b \, d^2\vv{x}$, where $b = \partial_x a_y - \partial_y a_x$ is the emergent ``magnetic'' field. Here $\wp_{132}$ denotes the closed loop $1 \rightarrow 3 \rightarrow 2 \rightarrow 1$ and $\Omega_{132}$ denotes the area enclosed by this path. Thus, $Q_{ijk} \approx 2i|C| \left( \sin(\phi_0) + \cos(\phi_0) \iint_{\Omega_{ijk}} b \, d^2\vv{x} \right) \approx 2i|C| \big( \sin(\phi_0) + \cos(\phi_0) \Omega_{ijk} \, b(\vv{r}_{ijk}) \big)$, where $C = |C| e^{i \phi_0}$ and $\vv{r}_{ijk}$ denotes the position of the three-site loop. \footnote{Since the photon momentum is small, we can ignore distances at the lattice scale. Thus, the precise definition of \protect{$\vv{r}_{ijk}$} do not concern us.} The $QQ$ correlator is thus converted into a $bb$ correlator. Hence:
\begin{widetext}
\begin{equation} \label{eq:QQ_correlator}
I_{A_{2g}}^{\txt{(gauge)}} \approx C' \int dt \: e^{i\Delta\omega t} \bra{i} 
	\left(\sum\nolimits_{\vv{R}} \sum\nolimits_{\tris} 
	b(\vv{r}_{ijk},t) e^{-i\vv{q}\cdot\vv{r}_{ijk}}\right)^\dagger
	\left(\sum\nolimits_{\vv{R}} \sum\nolimits_{\tris} 
	b(\vv{r}_{ijk},0) e^{-i\vv{q}\cdot\vv{r}_{ijk}} \right)
	\ket{i} \punct{,}
\end{equation}
\end{widetext}
where $C'$ is a numerical constant, and will not be kept track of below. Note that the photon momentum has been restored, which introduce the factor $e^{-i\vv{q}\cdot\vv{r}_{ijk}}$, with $\vv{q} = \vv{k}_i - \vv{k}_f$ the momentum transferred to the lattice. The $\sum_{\tris}$ is a shorthand for summing over the different three-site geometries on a unit cell with the appropriate coefficients, as is shown in Eq.~\ref{eq:A2g}.

Since $b = \partial_x a_y - \partial_y a_x$, $b(\vv{x}) = \sum_{k} (i \vv{k} \times \pol_k) a_{\vv{k}} e^{i\vv{k}\cdot \vv{x}} - (i \vv{k} \times \pol^*_k) a^\dg_{\vv{k}} e^{-i\vv{k}\cdot \vv{x}}$, where $\pol_k$ is the polarization of the emergent gauge field at momentum $\vv{k}$. Moreover, $a_k \ket{i} = 0$ and $\bra{i} a^\dg_k = 0$. Hence, upon Fourier transform,
\begin{equation}\label{eq:bb_correlator}
I_{A_{2g}}^{\txt{(gauge)}}  \propto	\bra{i} 
	(i \vv{q} \cdot \pol_q) a_{\vv{q}}(\Delta\omega) a^\dg_{\vv{q}}(0) (-i\vv{q}\cdot\pol^*_q) \ket{i} \punct{.}
\end{equation}

The correlator that we need to compute is thus one of a gauge field in 2+1 dimensions coupled to relativistic fermions (i.e., fermions described by the Dirac Hamiltonian Eq.~\ref{eq:H_Dirac}). This situation has been considered by Ioffe and Larkin\cite{Ioffe:PRB:1989} under the context of high-$T_c$ superconductivity, who found that:
\begin{equation} \label{eq:polarization}
\Pi_{\alpha\beta}(\vv{q}, \omega_E) = \frac{1}{8} \frac{\omega_E^2 \delta_{\alpha\beta} + v_F^2 q^2 \delta_{\alpha\beta} - v_F^2 q_\alpha q_\beta}{(\omega_E^2 + v_F^2 q^2)^{1/2}} \punct{,}
\end{equation}
where $\Pi_{\alpha\beta}(\vv{q}, \omega_E)$ is the polarization function of the gauge field in Euclidean spacetime, and $v_F = \chi J a/\sqrt{2}\hbar$ is the Fermi velocity at the Dirac node. Hence,
\begin{equation} \label{eq:spectral}
\begin{aligned}
I_{A_{2g}}^{\txt{(gauge)}} & \propto -q^2 \Imag 
	\left\{ \frac{8}{(v_F^2 q^2- \Delta\omega^2 + i \eta)^{1/2}}  \right\}
	+ \cdots \\
	& \propto \frac{q^2 \Theta(\Delta\omega - v_F q)}%
	{(\Delta\omega^2 - v_F^2 q^2)^{1/2}} + \cdots \punct{.}
\end{aligned}
\end{equation}

For herbertsmithite, $v_F$ is estimated\cite{Hermele:PRB:2008} to be $5.0\times 10^3$~m/s. Hence, even at back scattering and with optical light at wavelength $\lambda \approx$ 500~nm, $v_F q$ corresponds to a frequency shift of approximately $4.0$ cm$^{-1}$ only, which is too small to be resolved by current instruments. Therefore, the collective excitation associated with the gauge flux will appear as \emph{a characteristic $1/\omega$ singularity} in experiments. 

Physically, if the gauge boson is non-dissipative, the Green's function would have a simple pole, corresponding to a sharp delta-function-like signal. That we have a $1/\omega$ singularity in place of a delta function tells us that the gauge photon mode is strongly dissipative and is in fact overdamped. Following the analogy with the ordinary plasmon mode as stated above, this dissipative behavior of the emergent gauge boson can be thought of as the analog of the Landau damping.


\section{Discussion} \label{sect:discuss} 

In the previous sections we presented calculations of Raman intensity profile based on the Shastry--Shraiman formalism, assuming the validity of the U(1) Dirac spin-liquid state. We found a broad continuum in the Raman intensity profile in all symmetry channels, each displays a power-law behavior at low energy. Moreover, the profiles are found to be invariant under arbitrary rotations in the kagome plane. For the $E_g$ and the $A_{2g}$ channels, the continuum is accompanied by occasional sharp peaks that can be attributed to the various features of the DSL bands. In addition, the Raman intensity profile in the $A_{2g}$ channel also contains a characteristic $1/\omega$ singularity, which arose in our model from an excitation of the emergent U(1) gauge field.

However, several caveats in our theoretical predictions should be noted. First, in order to compare with experimental data, $\chi J$ must be converted to physical units. In Ref.~\onlinecite{Hermele:PRB:2008}, Hermele \etal\ estimated $\chi$ by fitting the spectrum of projected one-particle excitations to the mean-field band structure, and found that $\chi \approx 0.40$. Together with $J \approx 190 K$, this gives $\chi J \approx$ 56~cm$^{-1}$. However, there is considerable uncertainty in this estimation, and so it may be a good idea to take $\chi$ as a fitting parameter when comparisons with experiments are made.

Second, when the contribution to the Raman intensity profile by spinon-antispinon pairs are calculated, the excited spinons and antispinons have essentially been treated as free fermions. However, they should really be regarded as complicated composite fermions, which interact with each other through an effective gauge field. Consequently, the actual excitation spectrum of the quasiparticles will almost certainly look \emph{quantitatively} different from the ones presented here. Specifically, there may be finite lifetime effects, particularly prominent at high-energy, that causes the Raman intensity profile to be ``washed out'' compared with the ones presented here. Because of this, the sharp peaks that appear in Figs.~\ref{fig:Eg_corr} and \ref{fig:A2g} may not be present in the actual data. Furthermore, while the low-energy power-law structure of the Raman intensity profile is expected to survive, the detailed exponent is almost certainly modified from their mean-field values. Similarly, the contribution of Raman intensity in the $A_{2g}$ channel by the emergent gauge boson may scale as $I_{A_{2g}}^{\txt{(gauge)}} \propto (\Delta\omega)^{\alpha}$, with $\alpha$ modified from $-1$.

Third, in our derivation of the operators that correspond to the Raman transitions in the different channels, Eqs.~\ref{eq:Eg1}--\ref{eq:A2g}, we stopped at the zeroth order in $t/U$ and the leading order in $t/(\omega_i-U)$. While terms higher order in $t/U$ can be safely neglected, the same cannot be said for $t/(\omega_i-U)$, particularly near resonance ($\omega_i\approx U$). Therefore, these higher-order contributions, which modify the Raman intensity profile from those presented in the previous sections, may show up in actual data. In particular, the Raman intensity profile may not exhibit the abrupt drops as in Figs.~\ref{fig:Eg_corr} and \ref{fig:A1g}. Furthermore, since $\DOS_{\txt{1-pair}} \propto \Delta\omega$, a power-law with exponent much closer to $1$ may be found in the $E_g$ and $A_{1g}$ channels at low energy.

Fourth, while we argued that a $1/\omega$ singularity should be present in the $A_{2g}$ channel, we have lost track of the ratio between its contribution and that by the spinon--antispinon pairs. Since the intensity of this singularity is proportional $|\vv{q}|^2$, it may be difficult to detect in optical Raman spectroscopy, in which the momentum transfer $\vv{q}$ is much smaller than the inverse of lattice constant. 

In the introduction, we mentioned the VBS state as a close competitor of the DSL state, as well as the proposal by Cepas \etal\ to distinguish between the two using the angular dependence of the Raman intensity profile. In light of the results presented in this paper, there is another feature in the Raman spectrum that can be used to distinguish between the two states. For \emph{in general, in a VBS state the spin excitations is gapped, while in a spin-liquid state it is gapless. Consequently, the Raman intensity profile should show an exponential dependence in the former case, and a power-law dependence in the latter case.}

Recently, Wulferding and Lemmens \cite{Wulferding:unpub:2009} have obtained Raman intensity data for herbertsmithite. Their data, extending from 30 to 1500~cm$^{-1}$, shows a broad background that persists beyond 500~cm$^{-1}$, in addition to a quasielastic line and several sharp peaks at finite frequency shifts. Furthermore, at low temperature (5~K) the quasielastic line is suppressed and the low-energy portion of their data shows a linear dependence with respect to the Raman shift. While their data are quantitatively different from the results of our theoretical calculations presented in Secs.~\ref{sect:Eg}--\ref{sect:A2g}, the existence of a broad continuum can be seen as consistent with the U(1) Dirac spin-liquid model, even though the appearances of the other features would require the consideration of extra contributions (e.g., from the Zn impurities\cite{Vries:PRL:2008,Bert:JPConf:2009}) that are not present in our model.

\begin{acknowledgments}
We thank B. S. Shastry, B. I. Shraiman, and Naoto Nagaosa for discussions on the derivation of the Raman scattering $T$-matrix, Yi Zhou for discussions on the numerical calculations, and Peter Lemmens for discussions on his experimental data and for helpful comments on our manuscript. This research is partially supported by NSF under Grant No.~DMR-0804040.
\end{acknowledgments}

\appendix

\section{Derivation of the Raman transition in the \texorpdfstring{$A_{2g}$}{A2g} channel} \label{sect:A2g-derive}

In this section we shall consider the derivation of Raman transition rate in the $\bar{e}^{x}_{f} e^{y}_{i} - \bar{e}^{y}_{f} e^{x}_{i}$ channel in more details. For completeness, we shall present derivations not only for the kagome lattice, but also for the square, the triangular, and the hexagonal ones. Although the irreducible representation that corresponds to the polarization $\bar{e}^{x}_{f} e^{y}_{i} - \bar{e}^{y}_{f} e^{x}_{i}$ may be named differently in these lattices, we shall abuse notation and continue to refer to them as the ``$A_{2g}$'' channel.

We shall show that, contrary to claim by Shastry and Shraiman, the matrix elements in the $A_{2g}$ channel vanishes up to the $t^4/(\omega_i - U)^3$ order in the square lattice according to their formalism.

To extract the $A_{2g}$ channel from the general polarization matrix, note that given any particular hopping pathway, a ``reversed pathway'' can be constructed, in which all electron operators are conjugated and their order reversed [for example, $\cc{1}{2} \cc{2}{3} \cc{3}{1}$ is the reversed pathway of $\cc{1}{3} \cc{3}{2} \cc{2}{1}$]. Then, $\bar{e}^{x}_{f} e^{y}_{i} \mapsto \bar{e}^{y}_{f} e^{x}_{i}$ and the order the spin operators thus obtained are inverted. Hence, to the $t^4/(\omega_i - U)^3$ order, which corresponds to at most four spin operators, the only terms that survive in the $A_{2g}$ channel are the spin-chirality operators $\vv{S}_i \cdot (\vv{S}_j \times \vv{S}_k)$. Thus, in our derivation it suffices to extract the spin-chirality contributions from pathways whose initial and final currents are not co-linear.

To depict the hopping pathways efficiently, the following abbreviations are introduced in the diagrams. A thick (blue) arrow is used to indicate the initial or the final hop in which a holon--doublon pair is created or destroyed. For the internal hops, the movement of a doublon is indicated by a thin (magenta) unbroken arrow and the movement of a holon is indicated by a thin (magenta) broken arrow. Lower case roman letters are used to indicate the ordering of hops. Note that in this scheme, a solid magenta arrow from $i$ to $j$ corresponds to the electron operators $(c^\dg_j c_i)$, while a broken magenta arrow from $i$ to $j$ corresponds to the electron operators $(c^\dg_i c_j)$. 

The lowest order at which the spin-chirality term can show up is $t^3/(\omega_i - U)^2$, which corresponds to pathways with one internal hop. Such pathway can be found in the triangular or the kagome lattice, or when next-nearest hopping is included. We shall show that the contributions to the $A_{2g}$ channel by these pathways cancel in pairs at this order.

\begin{figure}[ht]
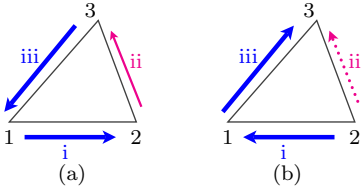

\begin{center}
\subfigure[\label{fig:1hop_e_short}]{\ehopshort} \qquad
\subfigure[\label{fig:1hop_h_short}]{\hhopshort}
\caption{\label{fig:1hop_short} Two types of one-internal-hop pathways. Thick (blue) arrows denote initial or final hops in which a holon--doublon pair is created or destroyed, thin (magenta) unbroken arrows denote the movement of doublons, and thin (magenta) broken arrows denote the movement of holons. Lower case roman letters are used to indicate the order of hops. In (a) the internal hop is performed by the doublon while in (b) the internal hop is performed by the holon.}
\end{center}
\end{figure}

It is easy to see that there are two types one-internal-hop pathways in general, both involving three lattice sites. In a pathway of the first type, a holon-doublon pair is created across a bond by an incident photon. Then, the doublon moves to a third site before recombining with the holon to emit a Raman-shifted photon [Fig.~\ref{fig:1hop_e_short}]. A pathway of the second type is similar, except that it is the \emph{holon} that moves to a third site before recombining [Fig.~\ref{fig:1hop_h_short}].

Applying the procedures as explained in Sec.~\ref{sect:SS_formulation}, the operator that corresponds to the pathway in Fig.~\ref{fig:1hop_e_short} is given by:
\begin{equation} \label{eq:1hop_e}
\begin{aligned}
T_{1,e} & = (\bar{\vv{e}}_f \cdot \vv{x}_{13})(\vv{e}_i \cdot \vv{x}_{21})
	\frac{(-i t_{13}) (-t_{32}) (-i t_{21})}{(\omega_i - U)^2}  \\
	& \qquad \times \cc{1}{3} \cc{3}{2} \cc{2}{1}\\
& = (\bar{\vv{e}}_f \cdot \vv{x}_{13})(\vv{e}_i \cdot \vv{x}_{21})
	\frac{t_{13} t_{32} t_{21}}{(\omega_i - U)^2} 
	\tr\left\{ \chi_3 \chi_2 \tchi_1	\right\} \\
& \doteq (\bar{\vv{e}}_f \cdot \vv{x}_{13})(\vv{e}_i \cdot \vv{x}_{21})
	\frac{t_{13} t_{32} t_{21}}{(\omega_i - U)^2}
	2 i \, \vv{S}_3 \cdot \vv{S}_2 \times \vv{S}_1 \punct{,}
\end{aligned}
\end{equation}
where  $\vv{x}_{ij} = \vv{x}_i - \vv{x}_j$ is the vector from site $j$ to site $i$, and ``$\doteq$'' denotes equality upon neglecting terms that do not contribute to the $A_{2g}$ channel. 

Similarly, the operator that corresponds to the pathway in Fig.~\ref{fig:1hop_h_short} is given by:
\begin{equation} \label{eq:1hop_h}
\begin{aligned}
T_{1,h} & = (\bar{\vv{e}}_f \cdot \vv{x}_{31})(\vv{e}_i \cdot \vv{x}_{12})
	\frac{(-i t_{31}) (-t_{23}) (-i t_{12})}{(\omega_i - U)^2} \\
	& \qquad \times \cc{3}{1} \cc{2}{3} \cc{1}{2} \\
& = (\bar{\vv{e}}_f \cdot \vv{x}_{31})(\vv{e}_i \cdot \vv{x}_{12})
	\frac{t_{31} t_{23} t_{12}}{(\omega_i - U)^2} 
	(-1) \tr\left\{ \chi_1 \tilde{\chi_2} \tchi_3	\right\} \\
& \doteq (\bar{\vv{e}}_f \cdot \vv{x}_{31})(\vv{e}_i \cdot \vv{x}_{12})
	\frac{t_{31} t_{23} t_{12}}{(\omega_i - U)^2}
	2 i \, \vv{S}_1 \cdot \vv{S}_2 \times \vv{S}_3 \\
& \doteq - T_{1,e} \punct{,}
\end{aligned}
\end{equation}
where $t_{ij}$ are assumed to be real in the last step.

Since the two pathways depicted in Fig.~\ref{fig:1hop_short} always come in pair, the contribution to the $A_{2g}$ channel by one-internal-hop pathways vanishes upon summing as claimed.

Now consider pathways that involve two internal hops, starting with the square lattice. Henceforth we shall assume that hopping is between nearest neighbors only, uniform, and real. The abbreviations $C_2 = t^4/(\omega_i-U)^3$ and $\SSS{i}{j}{k} = \vv{S}_i \cdot (\vv{S}_j \times \vv{S}_k)$ will also be used.

To count the two-internal-hop pathways in the square lattice systemically, we fix the initial holon at site 1 and the initial doublon at site 2, and align the coordinates so that $y = 0$ for site 1 and 2 and that $\vv{x}_{21} = \uv{x}$. All other pathways are clearly related to the ones satisfying the above conditions via symmetries. For the final hop and the initial hop to be non-collinear, a third site not collinear with site 1 and 2 must be involved, and we may further restrict our attention to pathways in which the third site lies in the $y > 0$ half-plane, since the remaining pathways are related to these via the mirror reflection $y \rightarrow -y$.


There are four pathways that satisfy the above restrictions, which are precisely the ones depicted in in Fig.~\ref{fig:sq_intro}. Applying the procedures as explained Sec.~\ref{sect:SS_formulation}, the contributions by these pathways are given by:
\begin{equation} \label{eq:process_a}
\begin{aligned}
T_{2,a} & = C_2 e_i^x (- \bar{e}_f^y) \cc{1}{3} \cc{3}{4} \cc{4}{2} \cc{2}{1} \\
	& = - C_2 e_i^x \bar{e}_f^y \tr\{ \chi_3 \chi_4 \chi_2 \tchi_1 \} \\
	& \doteq - i C_2 e_i^x \bar{e}_f^y (\SSS{3}{4}{1} + \SSS{3}{2}{1} + \SSS{4}{2}{1} - \SSS{3}{4}{2}) \punct{,}
\end{aligned} \end{equation}
\begin{equation} \label{eq:process_b}
\begin{aligned}
T_{2,b} & = C_2 e_i^x \bar{e}_f^y \cc{4}{2} \cc{3}{4} \cc{1}{3} \cc{2}{1}  \\
	& = C_2 e_i^x \bar{e}_f^y \tr\{ \chi_2 \tchi_1 \tchi_3 \tchi_4 \} \\
	& \doteq i C_2 e_i^x \bar{e}_f^y (\SSS{1}{3}{4} - \SSS{2}{3}{4} - \SSS{4}{2}{1} - \SSS{2}{1}{3}) \punct{,}
\end{aligned} \end{equation}
\begin{equation} \label{eq:process_c}
\begin{aligned}
T_{2,c} & = C_2 e_i^x (- \bar{e}_f^y) \cc{2}{4} \cc{1}{2} \cc{4}{2} \cc{2}{1} \\
	& = - C_2 e_i^x \bar{e}_f^y \Big((c_4 c^\dg_1) \cc{4}{2} \cc{2}{1} 
		+ (c_4 c^\dg_4) \cc{1}{2} \cc{2}{1} \Big)\\
	& = - C_2 e_i^x \bar{e}_f^y \Big( \tr\{(-1) \chi_4 \chi_2 \tchi_1 \} 
		+ \tr\{\chi_4\} \tr\{ \chi_2 \tchi_1 \} \Big) \\
	& \doteq - 2 i C_2 i e_i^x \bar{e}_f^y \SSS{4}{2}{1} \punct{,}
\end{aligned} \end{equation}
\begin{equation} \label{eq:process_d}
\begin{aligned}
T_{2,d} & = C_2 e_i^x \bar{e}_f^y \cc{3}{1} \cc{1}{2} \cc{1}{3} \cc{2}{1} \\
	& = C_2 e_i^x \bar{e}_f^y \Big( \cc{3}{2} \cc{1}{3} \cc{2}{1} 
		+ \cc{3}{3} \cc{1}{2} \cc{2}{1} \Big) \\
	& = C_2 e_i^x \bar{e}_f^y \Big( \tr\{(-1) \tchi_3 \chi_2 \tchi_1\} 
		+ \tr\{\tchi_3\} \tr\{\chi_2 \tchi_1\} \Big)\\
	& \doteq 2 i C_2 i e_i^x \bar{e}_f^y \SSS{3}{2}{1} \punct{.}
\end{aligned} \end{equation}

Summing all four terms, we found that $T_{2,a} + T_{2,b} + T_{2,c} + T_{2,d} \doteq 0$. Hence, for the square lattice with only nearest-neighbor hopping, the operator that corresponds to the Raman transition in the $A_{2g}$ channel, $O_{A_{2g}}$, vanishes to the $t^4/(\omega_i-U)^3$ order.

\begin{figure}[ht]
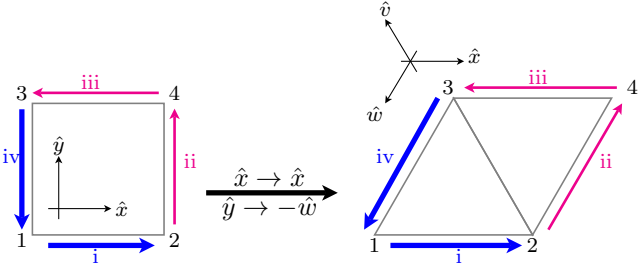

\begin{center}
\squaretriangular
\caption{\label{fig:sq-tri} Mapping between the pathways in the square lattice and the pathways in the triangular lattice.}
\end{center}
\end{figure}

Notice that the orthogonality between the $\uv{x}$ and $\uv{y}$ has not been invoked in the above derivation. Consequently, the above derivation carries to the triangular lattice upon mapping $\uv{x}$ and $\uv{y}$ in the square lattice to any two of bond directions in the triangular lattice. See Fig.~\ref{fig:sq-tri} for illustration. It can be checked that all two-internal-hop pathways with non-collinear initial and final hops in the triangular lattice can be obtained from such mappings and that there is no issue of double-counting. Hence, we conclude that $O_{A_{2g}}$ vanishes up to the $t^4/(\omega_i-U)^3$ order in the triangular lattice also.

Evidently, the criterion that any three non-collinear nearest-neighbor sites belong to a four-site loop is a crucial ingredient for the cancellations of the two-internal-hop contributions as seen above. This criterion is not met in the honeycomb lattice or in the kagome lattice. Hence, $O_{A_{2g}}$ may not vanish in these lattices at the $t^4/(\omega_i-U)^3$ order.

\begin{figure}
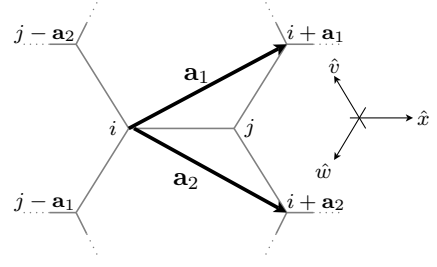

\begin{center}
\honeycomb
\caption{\label{fig:hex} The honeycomb lattice (thin gray lines), wherein the site label $i$, $j$, the unit vectors $\uv{x}$, $\uv{v}$, $\uv{w}$ (thin black arrows), and the primitive lattice vector (thick black arrows), are defined.}
\end{center}
\end{figure}

\begin{figure}
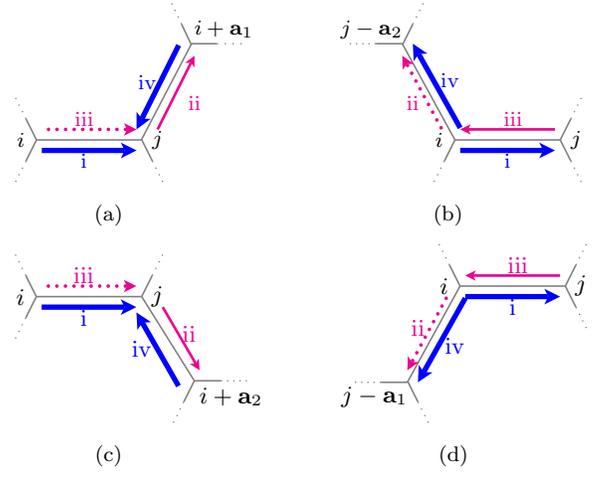

\begin{center}
\subfigure[\label{fig:hex_1}]{\honeycombone} \qquad 
\subfigure[\label{fig:hex_2}]{\honeycombtwo} \\
\subfigure[\label{fig:hex_3}]{\honeycombthree} \qquad
\subfigure[\label{fig:hex_4}]{\honeycombfour}
\caption{\label{fig:hex_2hop} Pathways with two internal hops that contribute to the $A_{2g}$ channel in the honeycomb lattice, with the initial holon fixed at site $i$ and initial doublon fixed at site $j$.}
\end{center}
\end{figure}

First consider the honeycomb lattice, which is shown in Fig.~\ref{fig:hex}, wherein the site labels $i$, $j$, the unit vectors $\uv{x}$, $\uv{v}$, $\uv{w}$ along bond directions, and the primitive lattice vectors $\vv{a}_1$, $\vv{a}_2$, are defined. For an initial holon at $i$ and an initial doublon at $j$, there are four two-internal-hop pathways, listed in Fig.~\ref{fig:hex_2hop}. Summing up their contributions, we get:
\begin{equation} \label{eq:hex_1x}
\begin{aligned}
T_{2,i,j} & \doteq 2 i C_2 \Big( 
	- e_i^x \bar{e}_f^w \SSS{i+\aone}{j}{i} 
	+ e_i^x \bar{e}_f^v \SSS{j-\atwo}{j}{i} \\ & \quad
	- e_i^x \bar{e}_f^v \SSS{i+\atwo}{j}{i}
	+ e_i^x \bar{e}_f^w \SSS{j-\aone}{j}{i} \Big) \punct{.}
\end{aligned}
\end{equation}

If the initial doublon is fixed at $j - \vv{a}_2$ or $j - \vv{a}_1$ instead, the contributions are, respectively,
\begin{equation} \label{eq:hex_1v}
\begin{aligned}
T_{2,i,j-\atwo} & \doteq 2 i C_2 \Big( 
	- e_i^v \bar{e}_f^x \SSS{i-\atwo}{j-\atwo}{i} 
	+ e_i^v \bar{e}_f^w \SSS{j-\aone}{j-\atwo}{i} \\ & \quad
	- e_i^v \bar{e}_f^w \SSS{i+\aone-\atwo}{j-\atwo}{i}
	+ e_i^v \bar{e}_f^x \SSS{j}{j-\atwo}{i} \Big) \punct{,}
\end{aligned}
\end{equation}
\begin{equation} \label{eq:hex_1w}
\begin{aligned}
T_{2,i,j-\aone} & \doteq 2 i C_2 \Big( 
	- e_i^w \bar{e}_f^v \SSS{i+\atwo-\aone}{j-\aone}{i} 
	+ e_i^w \bar{e}_f^x \SSS{j}{j-\aone}{i} \\ & \quad
	- e_i^w \bar{e}_f^x \SSS{i-\aone}{j-\aone}{i}
	+ e_i^w \bar{e}_f^v \SSS{j-\atwo}{j-\aone}{i} \Big) \punct{.}
\end{aligned}
\end{equation}

And the analog of Eqs.~\ref{eq:hex_1x}--\ref{eq:hex_1w}, when the holon is fixed at $j$, are given by:
\begin{equation} \label{eq:hex_2x}
\begin{aligned}
T_{2,j,i} & \doteq 2 i C_2 \Big( 
	- e_i^x \bar{e}_f^w \SSS{j-\aone}{j}{i} 
	+ e_i^x \bar{e}_f^v \SSS{i+\atwo}{j}{i} \\ & \quad
	- e_i^x \bar{e}_f^v \SSS{j-\atwo}{j}{i}
	+ e_i^x \bar{e}_f^w \SSS{i+\aone}{j}{i} \Big) \punct{,}
\end{aligned}
\end{equation}
\begin{equation} \label{eq:hex_2v}
\begin{aligned}
T_{2,j,i+\atwo} & \doteq 2 i C_2 \Big( 
	- e_i^v \bar{e}_f^x \SSS{j+\atwo}{i+\atwo}{j} 
	+ e_i^v \bar{e}_f^w \SSS{i+\aone}{i+\atwo}{j} \\ & \quad
	- e_i^v \bar{e}_f^w \SSS{j+\atwo-\aone}{i+\atwo}{j}
	+ e_i^v \bar{e}_f^x \SSS{i}{i+\atwo}{j} \Big) \punct{,}
\end{aligned}
\end{equation}
\begin{equation} \label{eq:hex_2w}
\begin{aligned}
T_{2,j,i+\aone} & \doteq 2 i C_2 \Big( 
	- e_i^w \bar{e}_f^v \SSS{j+\aone-\atwo}{i+\aone}{j} 
	+ e_i^w \bar{e}_f^x \SSS{i}{i+\aone}{j} \\ & \quad
	- e_i^w \bar{e}_f^x \SSS{j+\aone}{i+\aone}{j}
	+ e_i^w \bar{e}_f^v \SSS{i+\atwo}{i+\aone}{j} \Big) \punct{.}
\end{aligned}
\end{equation}

Summing Eqs.~\ref{eq:hex_1x}--\ref{eq:hex_2w} over all lattice vectors $\set{\vv{R}}$, and reorganize slightly, we finally obtain:
\begin{equation} \label{eq:hex}
\begin{aligned}
T_{2,\textrm{hex}} & \doteq 4 i C_2 \sum_{R} \Big(
	(\SSS{j}{i}{i-\aone} + \SSS{i}{j}{i+\aone})
		(e_i^x \bar{e}_f^w - e_i^w \bar{e}_f^x) \\ & \quad
	+ (\SSS{i-\atwo}{i}{j} + \SSS{i+\atwo}{j}{i})
		(e_i^v \bar{e}_f^x - e_i^x \bar{e}_f^v)	\\ & \quad
	+ (\SSS{j-\aone}{i}{i-\atwo} + \SSS{i+\aone}{j}{i+\atwo})
		(e_i^w \bar{e}_f^v - e_i^v \bar{e}_f^w)
	\Big) \\
& = 4 i C_2 \sum_{R} \Bigg(
	\Big( \xxtow + \xtoww \Big)(e_i^x \bar{e}_f^w - e_i^w \bar{e}_f^x) \\
	& \quad + \Big( \vvtox + \vtoxx \Big) (e_i^v \bar{e}_f^x - e_i^x \bar{e}_f^v) \\
	& \quad	+ \Big( \wwtov + \wtovv \Big) (e_i^w \bar{e}_f^v - e_i^v \bar{e}_f^w) 
	\Bigg) \punct{,}
\end{aligned}
\end{equation}
where graphical symbols are introduced on the second equality to denote the spin-chirality operators. Note that even though the site labels are omitted in the symbols, upon the summation over lattice vectors $\set{\vv{R}}$ there is no ambiguity as to which spin-chirality operator a particular symbol is referring to.

Using $e^v = -\frac{1}{2} e^x + \frac{\sqrt{3}}{2} e^y$ and $e^w = -\frac{1}{2} e^x - \frac{\sqrt{3}}{2} e^y$, Eq.~\ref{eq:hex} can be converted back to the Cartesian coordinates, which yields:
\begin{equation}\label{eq:hex_xy}
\begin{aligned}
T_{2,\textrm{hex}} & \doteq 2 \sqrt{3} i C_2 \sum_{R} 
	(e_i^y \bar{e}_f^x - e_i^x \bar{e}_f^y) \times \\
	& \nquad \Big(\wwtov \!+\! \wtovv \!+\! \vvtox \!+\! \vtoxx \!+\! \xxtow \!+\! \xtoww \Big) \punct{.}
\end{aligned}
\end{equation}

\begin{figure}
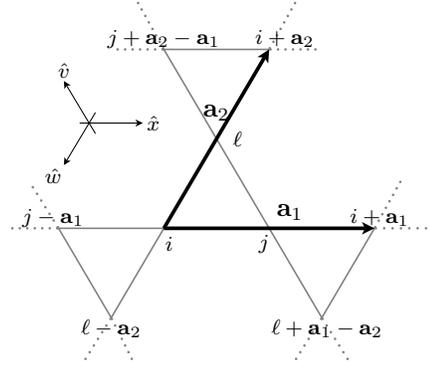

\begin{center}
\kagomesites
\caption{\label{fig:kagome} The kagome lattice (thin gray lines), wherein the site label $i$, $j$, the unit vectors $\uv{x}$, $\uv{v}$, $\uv{w}$ (thin black arrows), and the primitive lattice vector (thick black arrows), are defined.}
\end{center}
\end{figure}

\begin{figure}
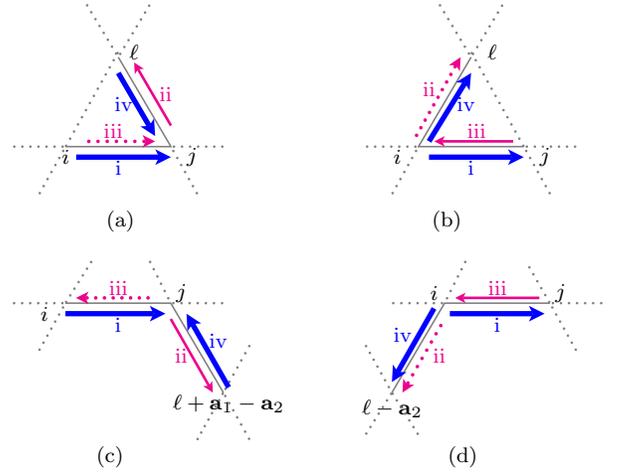

\begin{center}
\subfigure[\label{fig:kagome_1}]{\kagomeone} \qquad 
\subfigure[\label{fig:kagome_2}]{\kagometwo} \\
\subfigure[\label{fig:kagome_3}]{\kagomethree} \qquad
\subfigure[\label{fig:kagome_4}]{\kagomefour}
\caption{\label{fig:kagome_2hop} Pathways with two internal hops that contribute to the $A_{2g}$ channel in the kagome lattice, with the initial holon fixed at site $i$ and initial doublon fixed at site $j$.}
\end{center}
\end{figure}

Finally consider the kagome lattice, which is shown in Fig.~\ref{fig:kagome}, wherein the site labels $i$, $j$, $\ell$, unit vectors $\uv{x}$, $\uv{v}$, $\uv{w}$, and primitive lattice vectors $\vv{a}_1$, $\vv{a}_2$, are defined. Fixing the initial holon at site $i$ and the initial doublon at site $j$, four two-internal-hop pathways contribute to the $A_{2g}$ channel (Fig.~\ref{fig:kagome_2hop}). The sum of their contributions is:
\begin{equation} \label{eq:kag_1x}
\begin{aligned}
T_{2,i,j} & \doteq 2 i C_2 \Big( 
	+ e_i^x \bar{e}_f^v \SSS{\ell}{j}{i} 
	- e_i^x \bar{e}_f^w \SSS{\ell}{j}{i} \\ & \quad
	- e_i^x \bar{e}_f^v \SSS{\ell+\aone-\atwo}{j}{i}
	+ e_i^x \bar{e}_f^w \SSS{\ell-\atwo}{j}{i} \Big) \\
& = 2 i C_2 \Bigg( e_i^x \bar{e}_f^w \Big( \uptri +  \xxtow \Big) \\ & \quad
	- e_i^x \bar{e}_f^v \Big( \vtoxx + \uptri \Big)
	\Bigg) \punct{,}
\end{aligned}
\end{equation}
where graphical symbols are again introduced on the second equality. Note again that upon summing over all lattice vectors $\set{\vv{R}}$ there is no ambiguity as to which spin-chirality operator a particular symbol refers to.

By changing the site where the initial doublon is located, we get, upon summation, the following contribution to $O_{A_{2g}}$ by two-internal-hop pathways whose initial holon is located at site $i$:
\begin{widetext}
\begin{equation}\label{eq:kag_1}
\begin{aligned}
T_{2,i} & \doteq 2 i C_2 \Bigg(
	e_i^w \bar{e}_f^v \Big( \uptri \!+\! \dntri \!+\! \wwtov \!+\! \wtovv \Big) 
	- e_i^x \bar{e}_f^v \Big( \uptri \!+\! \dntri \!+\! \vtoxx \!+\! \vvtox \Big) 
	\\ & \quad
	+ (e_i^x \bar{e}_f^w - e_i^w \bar{e}_f^x)
		\Big( \uptri \!+\! \dntri \!+\! \xxtow \!+\! \xtoww \Big)
	\Bigg) \punct{.}
\end{aligned}
\end{equation}

Obtaining the contributions to $O_{A_{2g}}$ by two-internal-hop pathways whose initial holon is located at site $j$ or $\ell$ in an analogous manner, we finally get, upon summing over all lattice vectors and basis sites, 
\begin{equation}\label{eq:kag}
\begin{aligned}
T_{2,\txt{kag}} & \doteq 4 i C_2 \sum_{\vv{R}} \Bigg( 
	(e_i^x \bar{e}_f^w - e_i^w \bar{e}_f^x)
		\Big( \uptri \!+\! \dntri \!+\! \xxtow \!+\! \xtoww \Big) \\ & \nquad
	+ (e_i^v \bar{e}_f^x - e_i^x \bar{e}_f^v)
		\Big( \uptri \!+\! \dntri \!+\! \vtoxx \!+\! \vvtox \Big) 
	+ (e_i^w \bar{e}_f^v - e_i^v \bar{e}_f^w)
		\Big( \uptri \!+\! \dntri \!+\! \wwtov \!+\! \wtovv  \Big)
	\Bigg) \punct{.}
\end{aligned}
\end{equation}

Or, converting back to the Cartesian coordinates:
\begin{equation}\label{eq:kag_xy}
\begin{aligned}
T_{2,\txt{kag}} & \doteq 2 \sqrt{3} i C_2 \sum_{\vv{R}} 
	(e_i^y \bar{e}_f^x - e_i^x \bar{e}_f^y) 
	\Big( 3 \uptri + 3 \dntri + \wwtov + \wtovv 
	+ \vvtox + \vtoxx + \xxtow + \xtoww \Big) \punct{,}
\end{aligned}
\end{equation}
which is what we quoted in Eq.~\ref{eq:A2g}.
\end{widetext}

In summary, we found that, with only nearest-neighbor hopping, the $A_{2g}$ channel Raman $T$-matrix does not vanish in the honeycomb lattice or the kagome lattice at the $t^4/(U-\omega_i)^3$ order, but does so to this order in the square lattice and the triangular lattice.

\section{Derivation of \texorpdfstring{$O_{A_{1g}}$}{A1g operator} and \texorpdfstring{$O_{E_g}$}{Eg operator} to the \texorpdfstring{$t^4/(\omega_i-U)^3$}{t4/(w-U)3} order.} \label{sect:Eg-derive}

In this section we shall derive $O_{A_{1g}}$ and $O_{E_g}$ to the $t^4/(\omega_i-U)^3$ order for the kagome lattice. As already noted in Sec.~\ref{sect:SS_formulation}, at the $t^2/(\omega_i-U)$ order the Shastry--Shraiman formulation reproduces the Fleury--London Hamiltonian. In the $A_{1g}$ channel this gives rise to an operator proportional to the Heisenberg Hamiltonian, and in the $E_g$ channel it gives rise to the operators $O_{E_{g}^{(1)}}$ and $O_{E_{g}^{(2)}}$ as shown in Eqs.~\ref{eq:Eg1}--\ref{eq:Eg2}. 

At the $t^3/(\omega_i-U)^2$ order, it can be checked that the two types of pathways depicted in Fig.~\ref{fig:1hop_short} cancel each other not only in the $A_{2g}$ channel but also in the $E_g$ and $A_{1g}$ channels. Thus, it remains to consider 
pathways having two internal hops, which contributes at the $t^4/(\omega_i-U)^3$ order.

\begin{figure*}
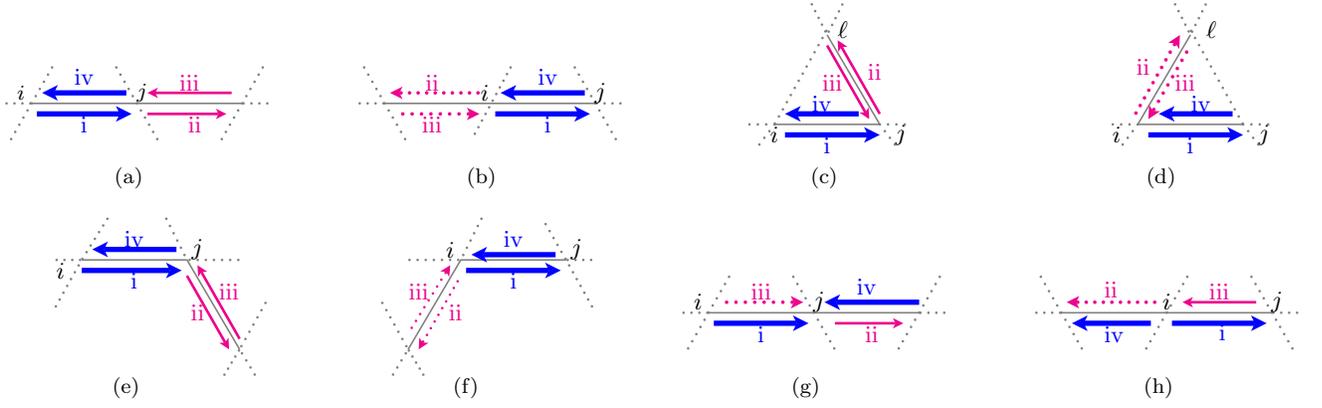

\begin{center}
\subfigure[\label{fig:kagome_05}]{\kagomefive} \qquad 
\subfigure[\label{fig:kagome_06}]{\kagomesix} \qquad
\subfigure[\label{fig:kagome_07}]{\kagomeseven} \qquad
\subfigure[\label{fig:kagome_08}]{\kagomeeight} \\
\subfigure[\label{fig:kagome_09}]{\kagomenine} \qquad
\subfigure[\label{fig:kagome_10}]{\kagometen} \qquad
\subfigure[\label{fig:kagome_11}]{\kagomeeleven} \qquad
\subfigure[\label{fig:kagome_12}]{\kagometwelve}
\caption{\label{fig:kagome_2hop_more} Pathways with two internal hops in a kagome lattice, with the initial holon fixed at site $i$ and initial doublon fixed at site $j$.}
\end{center}
\end{figure*}

By considering pathways and their ``reserved'' counterparts, in which all electron operators are conjugated and their order inverted, it is shown in Appendix~\ref{sect:A2g-derive} that only spin-chirality operators contribute to the $A_{2g}$ channel. From the same construction, it can be seen that the spin-chirality operators \emph{do not} contribute to the $E_g$ and $A_{1g}$ channels. Moreover, we are interested in the inelastic, and hence non-constant, part of the Raman transition operators. Therefore, to determine the $t^4/(\omega_i-U)^3$ order terms in $O_{E_{g}}$ and $O_{A_{2g}}$, it suffices to extract the spin dot product terms for each process.

Fixing the initial holon at site $i$ and the initial doublon at site $j$, there are eight more two-internal-hop pathways that contribute to the $A_{1g}$ and $E_g$ channels in addition to the four depicted in Fig.~\ref{fig:kagome_2hop}. These are listed in Fig.~\ref{fig:kagome_2hop_more}.

Applying the procedures as explained in Sec.~\ref{sect:SS_formulation}, the spin dot products resulting from Figs.~\ref{fig:kagome_1}, \ref{fig:kagome_2}, \ref{fig:kagome_07}, and \ref{fig:kagome_08} are, respectively:
\begin{equation} \label{eq:kag_a}
\begin{aligned}
T_{\txt{kag},a} & = C_2 e_i^x (-\bar{e}_f^v) \Big( \tr\{(-1) \chi_\ell \chi_j \tchi_i \} 
		+ \tr\{\chi_\ell\} \tr\{ \chi_j \tchi_i \} \Big) \\
	& \ \acute{=}\ C_2 e_i^x (-\bar{e}_f^v) (\SdotS{\ell}{i} - \SdotS{i}{j} - \SdotS{j}{\ell}) \punct{,}
\end{aligned}\end{equation}
\begin{equation}\label{eq:kag_b}
\begin{aligned}
T_{\txt{kag},b} & = C_2 e_i^x (-\bar{e}_f^w) \Big( \tr\{(-1) \chi_\ell \chi_j \tchi_i \} 
		+ \tr\{\chi_\ell\} \tr\{ \chi_j \tchi_i \} \Big) \\
	& \ \acute{=}\ C_2 e_i^x (-\bar{e}_f^w) (\SdotS{j}{\ell} - \SdotS{i}{j} - \SdotS{\ell}{i}) \punct{,}
\end{aligned} \end{equation}
\begin{equation} \label{eq:kag_c}
\begin{aligned}
T_{\txt{kag},c} & = C_2 e_i^x (- \bar{e}_f^x) 
		\cc{i}{j} \cc{j}{\ell} \cc{\ell}{j} \cc{j}{i} \\
	& = C_2 e_i^x (- \bar{e}_f^x) \tr\{\tchi_\ell\} \tr\{\chi_j \tchi_i \} \\
	& \ \acute{=}\ - 2 C_2 e_i^x (- \bar{e}_f^x) \SdotS{i}{j} \punct{,}
\end{aligned}\end{equation}
\begin{equation}\label{eq:kag_d}
\begin{aligned}
T_{\txt{kag},d} & = C_2 e_i^x (- \bar{e}_f^x) 
		\cc{i}{j} \cc{\ell}{i} \cc{i}{\ell} \cc{j}{i} \\
	& = C_2 e_i^x (- \bar{e}_f^x) \tr\{\chi_\ell\} \tr\{\chi_j \tchi_i \} \\
	& \ \acute{=}\ - 2 C_2 e_i^x (- \bar{e}_f^x) \SdotS{i}{j} \punct{,}
\end{aligned} \end{equation}
where $\acute{=}$ denotes equality upon neglecting additive constants and spin-chirality terms. The contributions of other pathways in Figs.~\ref{fig:kagome_2hop} and \ref{fig:kagome_2hop_more} can be obtained by relabeling sites and vectors that appear in Eqs.~\ref{eq:kag_a}--\ref{eq:kag_d}.

The sum over all pathways in Figs.~\ref{fig:kagome_2hop} and \ref{fig:kagome_2hop_more} gives:
\begin{widetext}
\begin{equation} \label{kag_1x_ss}
\begin{aligned}
T_{\txt{kag},i,j} & \ \acute{=}\ C_2 \Big( e_i^x \bar{e}_f^x 
		\big(\SdotS{i}{j-\aone} + \SdotS{j}{i+\aone}+14 \SdotS{i}{j} 
		- \SdotS{i}{i+\aone} - \SdotS{j}{j-\aone}\big) \\ & \quad 
	+ e_i^x \bar{e}_f^v 
		\big(\SdotS{i}{\ell+\aone-\atwo} + \SdotS{j}{\ell}
		- \SdotS{i}{\ell} - \SdotS{j}{\ell+\aone-\atwo}\big)
	+ e_i^x \bar{e}_f^w
		\big(\SdotS{j}{\ell-\atwo} + \SdotS{i}{\ell}
		- \SdotS{j}{\ell} - \SdotS{i}{\ell-\atwo}\big) 
	\Big) \punct{.}
\end{aligned}
\end{equation}

The contributions when the initial holon and/or the initial doublon are located at other sites can be obtained by relabeling. The sum over the locations of the initial holon and the initial doublon yields:
\begin{equation} \label{eq:kag_ss}
\begin{aligned}
T_{\txt{kag}} & \ \acute{=}\ 2 C_2 \sum_{\{\vv{R}\}} \bigg( e_i^x \bar{e}_f^x 
	\big( 16 \SdotS{i}{j} + 16 \SdotS{j}{i+\aone} - 2 \SdotS{i}{i+\aone} - 2 \SdotS{j}{j-\aone} \big) \\ & \quad
	+ e_i^v \bar{e}_f^v  \big( 16 \SdotS{j}{\ell} + 16 \SdotS{\ell}{j+\atwo-\aone} - 2 \SdotS{j}{j+\atwo-\aone} - 2 \SdotS{\ell}{\ell+\atwo-\aone} \big) \\ & \quad
	+ e_i^w \bar{e}_f^w \big( 16 \SdotS{i}{\ell} + 16 \SdotS{\ell}{i+\atwo} - 2 \SdotS{i}{i+\atwo} - 2 \SdotS{\ell}{\ell+\atwo} \big) \\ & \quad
	+ (e_i^x \bar{e}_f^w + e_i^w \bar{e}_f^x) 
		\big( \SdotS{j}{\ell-\atwo} + \SdotS{\ell}{j-\aone} - \SdotS{j}{\ell} - \SdotS{\ell-\atwo}{j-\aone} \big) \\ & \quad
	+ (e_i^x \bar{e}_f^v + e_i^v \bar{e}_f^x) 
	\big( \SdotS{i}{\ell+\aone-\atwo} + \SdotS{\ell}{j+\aone} - \SdotS{i}{\ell} - \SdotS{\ell+\aone-\atwo}{j+\aone} \big) \\ & \quad
	+ (e_i^v \bar{e}_f^w + e_i^w \bar{e}_f^v) 
	\big( \SdotS{i}{j+\atwo-\aone} + \SdotS{j}{i+\atwo} - \SdotS{i}{j} - \SdotS{j+\atwo-\aone}{i+\atwo} \big)
	\bigg) \punct{.}
\end{aligned}
\end{equation}
Projecting onto the $A_{1g}$ channel and neglecting a piece proportional to the Heisenberg Hamiltonian yields Eq.~\ref{eq:A1g}, while projecting into the $E_g$ channels yields the following contributions to $O_{E_g^{(1)}}$ and $O_{E_g^{(2)}}$ at the $t^4/(\omega_i-U)^3$ order, respectively:
\begin{align} 
\delta O_{E_g^{(1)}} & = C_2 \Big( (2 \SdotS{i}{j+\atwo-\aone} + 2 \SdotS{j}{i+\atwo} - \SdotS{j}{\ell-\atwo} - \SdotS{\ell}{j-\aone} - \SdotS{i}{\ell+\aone-\atwo} - \SdotS{\ell}{j+\aone} ) \notag \\ & \quad 
	- (2 \SdotS{i}{i+\aone} + 2 \SdotS{j}{j-\aone} - \SdotS{i}{i+\atwo} - \SdotS{\ell}{\ell+\atwo} - \SdotS{j}{j+\atwo-\aone} - \SdotS{\ell}{\ell+\atwo-\aone} ) \notag \\ & \quad 
	+ 7 (2 \SdotS{i}{j} + 2 \SdotS{j}{i+\aone} - \SdotS{i}{\ell} - \SdotS{\ell}{i+\atwo} - \SdotS{j}{\ell} - \SdotS{\ell}{j+\atwo-\aone}) \Big) \punct{,} \label{eq:Eg1_t4} \\
\delta O_{E_g^{(2)}} & = \sqrt{3} C_2 \Big( (\SdotS{j}{\ell-\atwo} + \SdotS{\ell}{j-\aone} - \SdotS{i}{\ell+\aone-\atwo} - \SdotS{\ell}{j+\aone} ) \notag \\ & \quad
	- (\SdotS{i}{i+\atwo} + \SdotS{\ell}{\ell+\atwo} - \SdotS{j}{j+\atwo-\aone} - \SdotS{\ell}{\ell+\atwo-\aone} ) \notag \\ & \quad
	+ 7 (\SdotS{i}{\ell} + \SdotS{\ell}{i+\atwo} - \SdotS{j}{\ell} - \SdotS{\ell}{j+\atwo-\aone}) \Big) \punct{.} \label{eq:Eg2_t4}
\end{align}
\end{widetext}

As explained in Sec.~\ref{sect:Eg}, it can be checked that $\delta O_{E_g^{(1)}}$ and $\delta O_{E_g^{(2)}}$ in the above equations do not produce any transition in the Dirac Hamiltonian Eq.~\ref{eq:H_Dirac}.


\end{document}